\documentclass{gGAF2e}
\usepackage{graphicx,natbib,bm,url,color}
\newcommand{\EQ}{\begin{equation}}
\newcommand{\EN}{\end{equation}}
\newcommand{\EQA}{\begin{eqnarray}}
\newcommand{\ENA}{\end{eqnarray}}
\newcommand{\eq}[1]{(\ref{#1})}
\newcommand{\EEq}[1]{Equation~(\ref{#1})}
\newcommand{\Eq}[1]{equation~(\ref{#1})}
\newcommand{\Eqs}[2]{equations~(\ref{#1}) and~(\ref{#2})}

\newcommand{\Eqss}[2]{equations~(\ref{#1})--(\ref{#2})}

\newcommand{\Sec}[1]{section~\ref{#1}}

\newcommand{\Fig}[1]{figure~\ref{#1}}
\newcommand{\Figs}[2]{figures~\ref{#1} and \ref{#2}}

\newcommand{\Figp}[2]{figure~\ref{#1}({#2})}

\newcommand{\Tab}[1]{table~\ref{#1}}

\newcommand{\bra}[1]{\langle #1\rangle}

\newcommand{\meanrho}{\overline{\rho}}

\newcommand{\means}{\overline{s}}

\newcommand{\meanT}{\overline{T}}

\newcommand{\nnn}{\hat{\mbox{\boldmath $n$}} {}}

\newcommand{\zzz}{\hat{\mbox{\boldmath $z$}} {}}

\newcommand{\FF}{\bm{F}}

\newcommand{\UU}{\bm{U}}

\newcommand{\uu}{\bm{u}}

\newcommand{\kk}{\bm{k}}

\newcommand{\xx}{\bm{x}}

\newcommand{\grav}{\mbox{\boldmath $g$} {}}
\newcommand{\nab}{\mbox{\boldmath $\nabla$} {}}
\newcommand{\nabad}{\nabla_{\rm ad}}
\newcommand{\nabrad}{\nabla_{\rm rad}}

\newcommand{\ttau}{\mbox{\boldmath $\tau$} {}}

\newcommand{\SSSS}{\mbox{\boldmath ${\sf S}$} {}}

\newcommand{\Rgas}{{\cal R}}

\newcommand{\ii}{{\rm i}}

\newcommand{\arccot}{{\rm arccot}  \, {}}
\newcommand{\DD}{{\rm D} {}}
\newcommand{\dd}{{\rm d} {}}
\newcommand{\const}{{\rm const}  {}}

\def\la{\mathrel{\mathchoice {\vcenter{\offinterlineskip\halign{\hfil
$\displaystyle##$\hfil\cr<\cr\sim\cr}}}
{\vcenter{\offinterlineskip\halign{\hfil$\textstyle##$\hfil\cr<\cr\sim\cr}}}
{\vcenter{\offinterlineskip\halign{\hfil$\scriptstyle##$\hfil\cr<\cr\sim\cr}}}
{\vcenter{\offinterlineskip\halign{\hfil$\scriptscriptstyle##$\hfil\cr<\cr\sim\cr}}}}}

\def\Bo{\mbox{\rm Bo}}

\def\Hp{H_{\!p}}
\def\Hr{H_{\!\rho}}

\def\cp{c_{\! p}}
\def\cv{c_{\! v}}
\def\cs{c_{\rm s}}
\def\csz{c_{\rm s0}}

\def\Teff{T_{\rm eff}}

\def\urms{u_{\rm rms}}

\def\cs{c_{\rm s}}

\def\sigmaSB{\sigma_{\rm SB}}

\def\half{{\textstyle{1\over2}}}
\def\threehalf{{\textstyle{3\over2}}}
\def\onethird{{\textstyle{1\over3}}}

\newcommand{\K}{\,{\rm K}}

\newcommand{\g}{\,{\rm g}}
\newcommand{\s}{\,{\rm s}}
\newcommand{\cm}{\,{\rm cm}}

\newcommand{\km}{\,{\rm km}}

\newcommand{\ks}{\,{\rm ks}}
\newcommand{\ms}{\,{\rm ms}}
\newcommand{\Mm}{\,{\rm Mm}}

\newcommand{\erg}{\,{\rm erg}}

\newcommand{\dgafd}[3]{, #2. {\em Geophys. Astrophys.\ Fluid Mech.} #1, doi:#3.}

\newcommand{\ypasj}[5]{, #5. {\em Publ. Astron. Soc. Jap.\ }#1, {\bf #2}, #3-#4.}

\newcommand{\ynaN}[4]{, #4. {\em New Astron.\ }#1, {\bf #2}, #3.}

\newcommand{\yana}[5]{, #5. {\em Astron.\ Astrophys.\ }#1, {\bf #2}, #3--#4.}

\newcommand{\yanaN}[4]{, #4. {\em Astron.\ Astrophys.\ }#1, {\bf #2}, #3.}

\newcommand{\ymn}[5]{, #5. {\em Month. Not. Roy.\ Astron.\ Soc.\ }#1, {\bf #2}, #3--#4.}

\newcommand{\ypr}[5]{, #5. {\em Phys.\ Rev.\ }#1, {\bf #2}, #3--#4.}

\newcommand{\yjcp}[5]{, #5. {\em J.\ Comp.\ Phys.\ }#1, {\bf #2}, #3--#4.}
\newcommand{\yjgr}[5]{, #5. {\em J.\ Geophys.\ Res.\ }#1, {\bf #2}, #3--#4.}

\newcommand{\yjas}[5]{, #5. {\em J.\ Atmosph.\ Sci.\ }#1, {\bf #2}, #3--#4.}

\newcommand{\yapj}[5]{, #5. {\em Astrophys.\ J.\ }#1, {\bf #2}, #3--#4.}
\newcommand{\yapjN}[4]{, #4. {\em Astrophys.\ J.\ }#1, {\bf #2}, #3.}
\newcommand{\yapjlN}[4]{, #4. {\em Astrophys.\ J.\ Lett.\ }#1, {\bf #2}, #3.}

\newcommand{\yapjS}[5]{, #5. {\em Astrophys.\ J.\ }#1, {\bf #2}, #3--#4}

\newcommand{\yapjs}[5]{, #5. {\em Astrophys.\ J.\ Suppl.\ }#1, {\bf #2}, #3--#4.}
\newcommand{\yapjl}[5]{, #5. {\em Astrophys.\ J.\ Lett.\ }#1, {\bf #2}, #3--#4.}

\newcommand{\ygafd}[5]{, #5. {\em Geophys.\ Astrophys.\ Fluid Dynam. }#1, {\bf #2}, #3--#4.}

\newcommand{\yjour}[6]{, #6. {\em #2} #1, {\bf #3}, #4--#5.}

\newcommand{\yproc}[7]{, #4. in {\em #5} (ed.\ #6), pp.\ #2--#3.\ #7.}
\newcommand{\ybook}[3]{ {\em #2}, #1 (#3).}

\graphicspath{{./fig/}{./png/}}

\begin{document}

\jvol{00} \jnum{00} \jyear{2018} 

\markboth{Axel Brandenburg and Upasana Das}{Time step constraint in radiation hydrodynamics}

\articletype{GAFD Special issue on ``Physics and Algorithms of the Pencil Code''}

\title{{\textit{The time step constraint in radiation hydrodynamics}$^\ast$
\thanks{$^\ast$Dedicated to Professor Ed A. Spiegel}}}

\author{AXEL BRANDENBURG$^{\rm a,b,c,d}$$^{\dag}$\thanks{$^\dag$Email: brandenb@nordita.org}
and UPASANA DAS$^{\rm a,b}$\vspace{6pt}
\\
$^{\rm a}$JILA, Box 440, University of Colorado, Boulder, CO 80303, USA\\
$^{\rm b}$Nordita, KTH Royal Institute of Technology and Stockholm University,\\
Roslagstullsbacken 23, SE-10691 Stockholm, Sweden\\
$^{\rm c}$Laboratory for Atmospheric and Space Physics,
University of Colorado, Boulder, CO 80303, USA\\
$^{\rm d}$Department of Astronomy, Stockholm University, SE-10691 Stockholm, Sweden
\\
\vspace{6pt}\received{\it \today,~ $ $Revision: 1.192 $ $} }


\maketitle

\begin{abstract}
Explicit radiation hydrodynamic simulations of the atmospheres of
massive stars and of convection in accretion discs around white dwarfs
suffer from prohibitively short time steps due to radiation.
This constraint is related to the cooling time rather than the radiative
pressure, which also becomes important in hot stars and discs.
We show that the radiative time step constraint is governed by the minimum of the
sum of the optically thick and thin contributions rather than the smaller
one of the two.
In simulations with the {\sc Pencil Code}, their weighting fractions
are found empirically.
In three-dimensional convective accretion disc simulations, the Deardorff
term is found to be the main contributor to the enthalpy flux
rather than the superadiabatic gradient.
We conclude with a discussion of how the radiative time step
problem could be mitigated in certain types of investigations.
\end{abstract}

\section{Introduction}

Numerical simulations have long played an essential role in
facilitating our understanding of hydrodynamic processes in astrophysics.
The cost of such simulations is determined not only by the numerical
mesh resolution, but also by the length of the time step.
In hydrodynamics with explicit time-stepping,
the maximum permissible time step decreases linearly
with increasing spatial mesh resolution in such a way that the information
that is passed from one time step to the next cannot propagate by more
than roughly one mesh spacing $\delta x$.
The length of the time step $\delta t$ is therefore of the order of
$\delta x/c$, where $c$ is the speed of the fastest propagating mode
(for example the speed of sound in subsonic compressible simulations).
This is known as the Courant-Friedrichs-Lewy (CFL) condition \citep{CFL28}
in ordinary hydrodynamics.
This condition is known to change for diffusive processes with 
diffusivity $\chi$. In radiative flows, the time taken to 
propagate information from one mesh point to the next is often estimated 
based on the diffusion approximation, which gives 
$\delta t \la C_{\rm diff} \delta x^2/\chi$ 
\citep[see, e.g.,][]{C04}, where $C_{\rm diff}$ is an empirical parameter,
which \cite{CK01} found to be $0.05$ in the context of magnetic diffusion.
Thus, decreasing the mesh width by a factor of two implies a reduction
of the time step by a factor of four.
Indeed, \cite{C04} quotes this as the main reason against explicit
radiation hydrodynamics in general.
While this may be true for certain cases, one must recall that 
the diffusion approximation is valid only in the
optically thick regime.

As one approaches the outer layers, the opacity decreases
sharply and the mean free path becomes long compared to other typical
scales in the system.
If one were to continue using the optically thick approximation, the
diffusivity would become excessively large.
This would have severe consequences for the length of the time step.
In reality, however, the optically thick approximation becomes invalid
and no stringent time step constraint is expected to occur in the optically
thin regime -- at least not for solar-type stars, although this may
change when more realistic opacities are invoked \citep{FSL12}.
However, empirically we know that for hotter stars
there can be layers in the proximity of
the photosphere where the time step constraint can become rather stringent.

In the present paper, we will be concerned with radiation transport using
what is nowadays often called long characteristics \citep{Mih78,Nor82}.
Relativistic effects are ignored and the radiation field
is propagated instantaneously across all rays without imposing any direct
time step constraint.
We refer to \cite{FOD09} for a treatment of time-dependent
radiative transfer simulations of cosmological reionisation using
long characteristics and to \cite{Pom79} for a discussion of the
non-equilibrium Marshak wave problem.
We also assume that the source function is just given by the Planck
function and thus ignore the possibility that it depends on the
mean intensity.
This implies that scattering is treated as true absorption, as is
commonly done \citep{FSL12}, but see the work of \cite{Ska00} for a
detailed treatment of scattering.
However, radiation interacts with the velocity field through radiation
pressure and the temperature field through heating and cooling processes.
While neither of these processes usually impose computationally
prohibitive time step constraints in solar physics \citep{SN89,SN98},
a serious time step constraint (more stringent than the hydrodynamic
time step) has been encountered empirically in numerical solutions of
hot atmospheres where radiation pressure contributes to the
hydrostatic equilibrium \citep{Spi06}.
One possibility is that the radiative pressure was responsible for the
empirically determined short time step.
More recently, the authors of the present work have encountered a similar time
step constraint when solving the radiation hydrodynamics equations for hot
accretion discs around white dwarfs.
However, since the radiation pressure was not included in those solutions,
it could not be held responsible here.

The fact that radiative time scales can be more restrictive
than hydrodynamic ones, and would hence lead to a stricter time step
constraint in explicit solvers, is not new \citep[see, e.g.,][]{C04,DSJ12}. 
However, there seem to be conflicting statements regarding this problem.
For example, \cite{DSJ12} derived a generalised CFL condition 
for the explicit radiation transfer solver in Athena
\citep{SGTHS08} using short characteristics.
However, instead of considering a general expression valid in 
both optically thick and thin regions, they approximated the radiative 
time step to be proportional to the time step due to 
the hydrodynamical CFL constraint locally in a grid zone
(see \Sec{RadiativeCoolingConstraint} for more details).  
This radiative time step is switched on whenever the optical depth 
per grid zone drops below unity anywhere in their computational domain.
Such an assumption does not seem justified, as will be explained below.
Furthermore, \cite{DSJ12} report on discrepancies and
problems with lower optical depths, as well as very short time steps 
when the radiation energy density dominates the gas energy density.
There is also a tendency to resort to semi-implicit
\citep[see, e.g.,][for the {\tt HERACLES} code]{GAH07}, or fully-implicit
radiation transfer solvers (see, e.g., \citealt{SN92a,SN92b,SMN92} for the Zeus code; 
see also \citealt{JSD12} for another module of Athena)
to bypass the radiative time step problem.
However, while implicit methods can avoid very small time steps,
they are computationally more expensive.
\cite{FSL12} proposed yet another time step constraint that is valid
both in optically thick and thin regimes.
However, no detailed study is presented.
Their proposal qualitatively agrees with ours, but is different in
that we suggest the additional presence of free parameters that
have not been suggested or motivated before, and that alleviate
the constraint in optically thin regions.

In view of the different and sometimes conflicting proposals
for the radiative time step in explicit radiation hydrodynamics,
there appears to be a need for a more a rigorous investigation.
Our hope is that by diagnosing in more detail the radiative time step
constraints in different situations, we would be in a better position
to avoid or mitigate the problem of short time steps.
One possibility might be to adopt certain changes in the model setup,
while still being able to capture the essential physics.
This will be discussed at the end of this paper.

\section{Radiative cooling constraint}
\label{RadiativeCoolingConstraint}

To quantify the expected time step constraint in radiation hydrodynamics,
we begin by computing the cooling time.
For that, we have to consider the radiation transport equation for
the intensity $I(\xx,t,\nnn)$, where $\xx$ is position, $t$ is time,
and $\nnn$ is the direction of the ray.
In the grey approximation, the radiation transport equation is
\EQ
\nnn \bm \cdot \nab I=-\kappa\rho\,(I-S),
\label{ngradI_eqn}
\EN
where $\rho$ is the density, $\kappa$ is the opacity per unit mass, and
$S(\xx,t)$ is the source function, which we will assume to be given
by the Planck function, i.e., $S=(\sigmaSB/\pi)T^4$, with $\sigmaSB$
being the Stefan-Boltzmann constant, and $T$ is the temperature.

To gain insight into the nature of radiation in the optically thin and
thick cases, it is useful to adopt a model where we can assume constant
coefficients, which allows us to use Fourier transformation.
We also adopt the Eddington approximation, where the moment expansion is
closed by assuming the radiation pressure to be isotropic and given by
$\onethird\delta_{ij}J$, where $J=\int_{4\pi}I\,\dd\Omega/4\pi$ is the
mean intensity and $\dd\Omega$ is the differential over the solid angle.
This yields \citep{Edw90}
\begin{equation}
\onethird(\ell\nab)^2J=J-S,
\end{equation}
where $\ell=(\kappa\rho)^{-1}$ is the photon mean-free path.
Note that $(\ell\nab)^2=\ell^2\nab^2$ only when $\ell$ is constant in space.
In the absence of any heating and cooling processes other than
the negative radiative flux divergence, $-{\nab \bm \cdot \FF_{\rm rad}}$,
which is proportional
to $(\ell\nab)^2J$, the temperature evolution is governed by the
radiative heat equation
\begin{equation}
\rho\cp{\DD T\over\DD t}-{\DD p\over\DD t}={\textstyle{4\pi\over3}}\,\kappa\rho\,(\ell\nab)^2J,
\label{opt_thickthin}
\end{equation}
which is valid both in the optically thick and thin cases.
Here, $p$ is the pressure, $\cp$ is the specific heat at constant pressure,
and $\DD/\DD t=\partial/\partial t+\uu \bm \cdot \nab$ is the
advective derivative.
For the purpose of the present discussion, we assume $p=\const$ and
omit the $\DD p/\DD t$ term.
Linearising \Eq{opt_thickthin} about a hydrostatic homogeneous
equilibrium solution with $\uu=\bm{0}$, $T=\const$, and $\rho=\const$,
and assuming the solution to be proportional to
$e^{\ii\kk \bm \cdot \xx-\lambda t}$, where $\kk$ is the wavevector,
we find for the cooling or decay rate $\lambda$ the expression
\citep{US66}
\begin{equation}
\lambda={c_\gamma\over\ell}\,{k^2\ell^2/3\over1+k^2\ell^2/3}
={c_\gamma k^2\ell/3\over1+k^2\ell^2/3}
={\chi k^2\over1+k^2\ell^2/3},
\label{US66expr}
\end{equation}
where $k=|\kk|$ is the wavenumber,
\begin{equation}
c_\gamma=16\sigmaSB T^3/\rho \cp
\label{cgam_def}
\end{equation}
is the characteristic velocity of photon diffusion
\citep{BB14}, and $\chi=c_\gamma\ell/3$ is the radiative diffusivity.
The quantity $c_\gamma$ is related to the radiative relaxation
time $\ell/c_\gamma$ (equivalent to $q^{-1}$ of \citealt{US66}).
It is smaller than the speed of light by roughly the ratio of
radiative to thermal energies.
Expression \eq{US66expr} has been
obtained under the Eddington approximation and deviates
only slightly from the exact expression obtained by \cite{Spi57}, which 
can be written as
\begin{equation}
\lambda_{\rm exact} = {c_\gamma\over\ell}\, \left(1 - 
\frac{1}{k\ell} \arccot \frac{1}{k\ell} \right)
={c_\gamma\over\ell}\, \left(1 - \frac{\arctan k\ell}{k\ell} \right)
\la\lambda.
\label{S57expr}
\end{equation}
The largest value of the ratio $\lambda/\lambda_{\rm exact}$ is 1.29 at $k\ell=2.53$.

The former expression \eq{US66expr} has the advantage that its inverse is
the sum of two terms, allowing us to easily analyse the different regimes.
When $\ell$ is small in the sense that $k\ell\ll1$, the cooling rate is
$\lambda\approx\chi k^2=c_\gamma\ell k^2/3$, which corresponds to the
usual expression in the optically thick limit.
On the other hand, in the optically thin limit,
when $\ell$ is large ($k\ell\gg1$), we have
$\lambda\approx c_\gamma/\ell$, so cooling becomes independent of $k$.
The resulting value of $\lambda$ is much smaller than the value of
$c_\gamma\ell k^2/3$, which it would be if one continued using the
expression for the optically thick case even though $k\ell\gg1$.
Thus, when using the correct expression, it may seem that 
radiation is less likely to be a limiting factor in the time step 
consideration. 
However, as we will show in this work, this is not necessarily true.

To arrive at an expression for the time step constraint due to radiation, 
we follow the idea of \cite{FSL12} and
express the cooling time as the inverse of the cooling rate,
$\tau_{\rm cool}=\lambda^{-1}$, where
\begin{equation}
\tau_{\rm cool}={1\over\chi k^2}+{\ell\over c_\gamma}.
\label{taucool}
\end{equation}
The first and second terms on the right-hand side of \eq{taucool}
characterise the contributions from the optically thick and thin parts,
respectively, and $k$ should be replaced by the Nyquist wavenumber,
$k_{\rm Ny}=\pi/\delta z$.
Here and in the following sections, we restrict our attention to the
vertical mesh spacing $\delta z$.
However, in more than one dimension, the $k^2$ factor gains additional
contributions $k_x^2=(\pi/\delta x)^2$ and $k_y^2=(\pi/\delta y)^2$ for
the mesh spacings $\delta x$ and $\delta y$ in the $x$ and $y$ directions.
Assuming $\delta x=\delta y=\delta z$, we would need to replace
$k^2\to k_x^2+k_y^2+k_z^2=3(\pi/\delta z)^2$, or, more generally for
$D$ dimensions, by $(\pi/\delta z)^2D$.
In general, the direction with the finest mesh spacing will impose
the strongest constraint.

The parts for the optically thick and thin regimes may contribute
with different non-dimensional prefactors to the actual
(empirically determined) time step constraint.
Nevertheless, we will assume that the time step limit still depends
on the sum of these two parts, just like in \Eq{taucool}, because
this ensures that in the optically thin regime, the much faster cooling
rate relevant to the optically thick limit does not contribute.
For a global simulation to be stable in both regimes, the time step can
therefore not exceed the shortest value of the sum anywhere in the domain.
Keeping this in mind, we define the radiative time step constraint to be
\begin{equation}
\delta t_{\rm rad}=C_{\rm rad}^{\rm thick}\,{\delta z^2\over\chi D}
+C_{\rm rad}^{\rm thin}{\ell\over c_\gamma},
\label{dtrad_new}
\end{equation}
such that the corresponding maximum permissible time step is given by
\begin{equation}
\delta t \leq \min(\delta t_{\rm rad}),
\label{dtrad}
\end{equation}
where $C_{\rm rad}^{\rm thick}$ and $C_{\rm rad}^{\rm thin}$ are 
dimensionless coefficients.
If we replace $k$ by the Nyquist wavenumber in $D$ dimensions,
as done above, we have $C_{\rm rad}^{\rm thick}=1/\pi^2$, 
To estimate the value of $C_{\rm rad}^{\rm thin}$, we must
consider the properties of the time stepping scheme.
If we used a first order Euler scheme, then
$\delta t\,c_\gamma/\ell<2$ is required for stability \citep{SB02}, i.e.,
$C_{\rm rad}^{\rm thin}<2$.
Since this factor is larger than unity, it also has implications for
the actual value of $C_{\rm rad}^{\rm thick}$, which would therefore
be closer to $C_{\rm rad}^{\rm thick}=2/\pi^2\approx0.2$.
If we were to combine the Euler time stepping scheme with a less
accurate second order discretisation of the Laplacian, the effective
Nyquist wavenumber would only be $2/\delta z$ instead of $\pi/\delta z$,
and one would obtain $C_{\rm rad}^{\rm thick}<2/2^2=0.5$.
This is a well known result from the von Neumann stability analysis
of the heat equation \citep{CN47,CFvN50}.

Note that $c_\gamma$ enters in both terms of \Eq{dtrad_new}. 
In the first term, it indeed plays the role of a photon 
diffusion velocity, but in the second term it represents a 
characteristic photon crossing velocity in an optically thin medium.

In direct contrast with the above calculation, we quote the radiative time 
step constraint used by \cite{DSJ12},
\begin{equation}
\delta t_{\rm rad}^{\rm Athena} \propto \min(\Bo) \,
\min (\delta z/\cs)  ~;
\label{eq_davis12}
\end{equation} 
see their equations (29) and (43), where $\Bo=16\,\cs/c_\gamma$
is the local Boltzmann number and $\cs$ is the sound speed.
This time step constraint is used in the explicit radiation
transfer module of Athena.
The time step is here explicitly proportional to the usual CFL condition.
Thus, \Eq{eq_davis12} does not properly account 
for the transition from the optically thick to thin regions, 
as mentioned in the introduction.
Note that the Boltzmann number is commonly 
invoked in radiative flows \citep[see, e.g.,][]{C04,DSJ12}, and it is a 
measure of when radiation becomes important in a problem.

\EEq{taucool} was also the basis of the time step constraint used
in the {\tt CO5BOLD} code \citep{FSL12}, which is used to model 
solar and stellar surface convection. In that code, radiation transport 
is also treated explicitly (although there are modules allowing for
semi-implicit or fully implicit treatments).
According to the {\tt CO5BOLD} user 
manual\footnote[1]{See \url{http://www.astro.uu.se/~bf/co5bold_main.html}.}, 
there is only {\em one} prefactor, which they refer to as the ``radiative
Courant" factor and they recommend it to be adjusted by trial and error.
In this work, we introduce two separate coefficients
$C_{\rm rad}^{\rm thick}$ and $C_{\rm rad}^{\rm thin}$,
which can be thought of as the radiative Courant-like
factors in the optically thick and thin regions, respectively.
Both coefficients depend on the time stepping scheme, but the first one
also depends on the spatial discretisation scheme.

It is useful to reflect again on the somewhat unusual form of \Eq{dtrad}
as a time step constraint, because the usual CFL condition is formulated
in terms of the shortest one of several constraints, e.g., $\delta
t=\min(\delta z/\cs,\,\delta z^2/\chi D)$.
Alternatively, one could express the inverse time step
as the sum of the inverse of the contributions.
In \eq{dtrad_new}, by contrast, $\delta t_{\rm rad}$ itself is determined
as a sum of two time steps.
In full radiation hydrodynamics, the maximum permissible time step
will therefore be the minimum of $\delta t_{\rm rad}$, as given by 
\Eq{dtrad_new},
and the usual CFL and viscous constraints, i.e.,
\begin{equation}
\delta t=\min\left(C_{\rm rad}^{\rm thick}\,{\delta z^2\over\chi D}
+C_{\rm rad}^{\rm thin}{\ell\over c_\gamma},\,
C_{\rm CFL}{\delta z\over\cs},\, C_{\rm visc}{\delta z^2\over\nu D}
\right).
\label{dtfull}
\end{equation}
For the {\sc Pencil Code} with its default third order time stepping scheme,
$C_{\rm CFL}=0.9$ is the usual CFL number, $C_{\rm visc}=0.25$ is the
viscous time step constraint, with $\nu$ being the kinematic viscosity and 
$D$ the dimensionality.
In our present one-dimensional calculation, we have $D=1$.
In three dimensions, $C_{\rm visc}/D=0.08$ is in fact slightly
larger than the value $C_{\rm diff}=0.05$ quoted by \cite{CK01}.
Again, the dimension $D$ enters because the discretised form of
the second derivative, relevant in the optically thick
formulation, has a larger coefficient at the center
point where the derivative is evaluated.\footnote[2]{For
a second order discretisation, for example, we have
$\nabla^2f_i=(f_{i+1}-2f_{i}+f_{i-1})/\delta x^2$ in one dimension, but
$\nabla^2f_{ij}=(f_{i+1\,j}+f_{i\,j+1}-4f_{ij}+f_{i-1\,j}+f_{i\,j-1})/
\delta x^2$ in two dimensions, so at the center point, $f_{ij}$, the
coefficient increases from $2/\delta x^2$ to $4/\delta x^2$ in two
dimensions, and to $6/\delta x^2$ in three dimensions.
This applies analogously also to the sixth order discretisation
used in the {\sc Pencil Code}, except that the coefficient now 
increases to $49/18\approx2.72$ instead of 2 per direction as 
for the second order case.
Also, for a given function $f$, the value of $-(\nabla^2 f)/f$
depends on the numerical scheme.
For a checkerboard pattern of $f$ (e.g., an alternating sequence with $-1$,
$+1$, etc, in one spatial dimension), using a second order scheme, the value
is $-4/\delta z^2$ per direction, while the analytic value is
$-\pi^2/\delta z^2\equiv k_{\rm Ny}^2$, which is more than twice as much.}

We emphasise that there is no analogy in how the optically thick and
thin contributions enter into the time step constraint and how the
usual CFL and viscous constraints enter.
This becomes strikingly clear by stating
\begin{equation}
\delta t\neq\min\left(C_{\rm rad}^{\rm thick}\,{\delta z^2\over\chi D},\,
C_{\rm rad}^{\rm thin}{\ell\over c_\gamma},\,
C_{\rm CFL}{\delta z\over\cs},\, C_{\rm visc}{\delta z^2\over\nu D}
\right).
\label{dtfull_wrong}
\end{equation}
Later in this paper, we will see examples where either $\delta z^2/\chi$
or $\ell/c_\gamma$ may be very small, and yet, neither of those affect the
time step if the other term is large.

The goal of the present paper is to test the validity of \eq{dtrad_new}
in the case when radiation is treated with long characteristics.
We also compare with the usual CFL condition, where the time step
is constrained by $\delta t_{\rm s}= C_{\rm CFL} \delta z/\cs$.
Since the role of $c_\gamma$ is not entirely clear, especially when
radiation pressure also enters the problem, we ask whether a 
similarly defined quantity $\delta t_\gamma=C_\gamma \delta z/c_\gamma$,
as suggested by the work of \cite{DSJ12},
might constrain the time step further, even though it does not 
explicitly feature in \Eq{dtfull}.
Our numerical experiments reported below show that 
$\delta t_\gamma$ itself does not constrain the time 
step, although the ratio $\cs/c_\gamma$ may still play an important role;
see discussion in \Sec{secMitigate}.
This can be better understood by recalling that 
the Boltzmann number $\Bo$ is proportional to $\cs/c_\gamma$.
Thus, a smaller value of $\Bo$ signifies that the energy transport 
is radiation dominated and vice versa.

If $c_\gamma$ were to enter the time step constraint directly through
a quantity $t_\gamma$, it would be natural to expect that $C_\gamma$
would be of the order of $C_{\rm CFL}$.
Hence, we use $C_\gamma=C_{\rm CFL}=0.9$ in our plots below.
Regarding the value of $C_{\rm rad}^{\rm thick}$, we expect it
to be comparable to $C_{\rm visc}=0.25$, but our experiments
reported below seem to be consistent with a slightly smaller
value of $C_{\rm rad}^{\rm thick}=0.2$, so we will use that
value in all the corresponding plots shown below.
Finally, regarding the value of $C_{\rm rad}^{\rm thin}$, it is important to
note that it enters without a $\delta z$ term; see second term in
\eq{dtrad_new}.
In order to obtain a preliminary estimate, we empirically test
the radiative time step constraint in the optically thin case by using a
one-dimensional model. In this way, we find that 
$C_{\rm rad}^{\rm thin}\approx4$ and we use this value in the 
plots shown below; see \Sec{Unstratified} for details. 
We find that the values of the various coefficients quoted here 
are consistent with the empirically determined ones, as discussed in 
detail in \Sec{DNSdiscmodels} and \Tab{Tsum}.

In this paper, we discuss two distinct models where severe time step
constraints have been encountered.
One is the model of \cite{Spi06} and the other is a local model of an
accretion disc, similar to that of \cite{CBHH18}.
Radiation pressure is included in the former, but not in the latter.
It will turn out that the time step constraints are quite different
from each other in the two cases, although this difference is not
explicitly linked to the presence or absence of radiation pressure.
Nevertheless, when the radiation pressure becomes extremely large,
it can in principal also restrict the time step.
We discuss this possibility at the end of
our penultimate section \ref{secMitigate}.

Finally, let us note that the radiation magnetohydrodynamic shearing-box 
simulations of accretion discs by \cite{CBHH18} were carried out
using the Zeus code. Time step problems were 
probably not encountered in this case, as Zeus employs an 
implicit radiation transfer solver.

\section{Our models}

\subsection{The basic equations}

We consider here the nonrelativistic radiation hydrodynamics
equations solved by default in the {\sc Pencil Code}. Note, however, 
that we work in what is referred to as the static diffusion limit.
Defining $\beta=u/c$, where $u$ is the characteristic velocity of the 
system and $c$ the speed of light, the static diffusion limit is valid
when $\beta \tau \ll 1$ in optically thick regions where
$\tau\equiv\int\rho\kappa\,\dd z \gg 1$;
see \cite{Kru07} for a detailed discussion of 
the various regimes in radiation hydrodynamics. 

The basic dependent variables are the logarithmic density $\ln\rho$,
the velocity $\uu$, and the specific entropy $s$, which obey the equations
\EQ
{\DD\ln\rho\over\DD t}=-\nab \bm \cdot \uu,
\label{dlnrho}
\EN
\EQ
\rho{\DD\uu\over\DD t}=-\nab p + \rho\grav + {\rho\kappa\over c}\FF_{\rm rad}
+\nab \bm \cdot\ttau,
\label{DuDt}
\EN
\EQ
\rho T{\DD s\over\DD t}={\cal H} - \nab \bm \cdot\FF_{\rm rad}
+\ttau\bm{:}\nab\UU,
\label{DsDt}
\EN
\EQ
\nnn \bm \cdot\nab I=-\kappa\rho\,(I-S),\quad
\FF_{\rm rad}=\int_{4\pi}\nnn I\,\dd\Omega,\quad
\nab \bm \cdot\FF_{\rm rad}=\int_{4\pi}(I-S)\,\dd\Omega,
\label{ngradI}
\EN
where $\grav=(0,0,-g)$ is the gravitational acceleration in
Cartesian coordinates $(x,y,z)$,
$\FF_{\rm rad}$ is the radiative flux,
$\ttau=2\rho\nu\SSSS$ is the stress tensor if there is just shear viscosity,
${\sf S}_{ij}=\half(\partial_i u_j+\partial_j u_i)-\onethird
\delta_{ij}\nab \bm \cdot\uu$ are the components of the traceless rate-of-strain tensor,
${\cal H}$ is a heating function to be specified below,
$\ttau\bm{:}\nab\UU\equiv\tau_{ij} \partial u_i/\partial x_j$
is the viscous heating term, and
\Eq{ngradI_eqn} has been restated in \eq{ngradI} for completeness.
The specific entropy is related to pressure and density via
$\DD s=\cv\DD\ln p-\cp\DD\ln\rho$, where $\cv$ is the specific heat
at constant volume.
In the model for the upper layers of a star, we have $g=\const$, while
in our accretion disc model with local Keplerian angular velocity $\varOmega$
(not to be confused with the solid angle $\Omega$),
$g=\varOmega^2z$ increases linearly with the distance $z$
from the midplane.
Its sign is such that gravity pulls toward the midplane from
above and below $z=0$.

The equations of radiation transport \eq{ngradI} have been implemented
into the {\sc Pencil Code} by \cite{HDNB06}.
This implementation was then used in the solar context \citep{HNSS07}
and later also in more idealised problems \citep{BB14}.
In the outer layers of the Sun, partial ionisation is also important,
so one needs to solve the Saha equation, for which temperature needs
to be known.
It is then advantageous to use $\ln T$ as the dependent variable instead
of $s$.
Again, this implementation into the {\sc Pencil Code} goes back to the
work of \citep{HNSS07}, and more idealised models with ionisation and
radiation have been considered by \cite{BB16}.

In the presence of shocks, it is often useful to increase the viscosity
locally in those regions where the velocity converges, i.e., where the
flow divergence is negative or $\nab \bm \cdot\uu<0$.
This approach goes back to \cite{vNR50}.
In practice, one defines the shock viscosity as
\begin{equation}
\nu_{\rm shock}=C_{\rm shock} \delta x^2 \bra{-\nab \bm \cdot\uu}_+,
\end{equation}
where $\bra{...}_+$ denotes a running five point average over all positive
arguments, 
$C_{\rm shock}$ is the dimensionless coefficient of the shock viscosity,
and $\delta x=\delta y=\delta z$ are the mesh spacings in all
three directions.
This shock viscosity is applied as a bulk viscosity, i.e., $\ttau$
in \Eqs{DuDt}{DsDt} is given by
\EQ
\tau_{ij}=2\rho\nu{\sf S}_{ij}+\rho\nu_{\rm shock}\delta_{ij}\nab \bm \cdot\uu.
\EN
At the end of this paper, we present one case where a shock viscosity
is applied and show how it affects the time step
(\Figs{ppdt_Strat_H1.0e-05_z20_nu5em4a4}{pslice_Strat_H1.0e-05_z20_nu5em4a4}
of \Sec{3DdiscDNS}).

\subsection{Numerical treatment within the {\sc Pencil Code}}
\label{RadiationTransport}

In the {\sc Pencil Code}, all derivatives are usually approximated by
sixth order finite differences; see \cite{Bra03} for details.
The third-order time stepping scheme of \cite{Wil80} is used.
Thus, during each of the three substeps, the right-hand side of the
equations is evaluated three times.
The code is not explicitly conservative, but mass, energy, and momentum
are conserved within the  discretisation error of the scheme, so we can
use their conservation properties to gauge the accuracy of the scheme.
Normally, when doubling the resolution, the error decreases by a factor
of $2^6=64$, as was demonstrated by \cite{BHB11}.

The radiation transport is solved using long characteristics.
A detailed account of its implementation has been given by \cite{HDNB06}.
For a brief description of the most important numerical aspects, we refer
the reader to section~2.4 of \cite{BB14}.
Instead of integrating along a geometric line segment $\dd l$, we
integrate \eq{ngradI_eqn} over optical depth $\dd\tau=-\kappa\rho\,\dd l$
for a number of rays with direction $\nnn$, such that $\dd I/\dd\tau=I-S$. 
In formulating the boundary conditions, we distinguish between two types of rays.
For rays that are perfectly horizontal, we assume periodicity.
For all other rays, we normally assume that no radiation enters
at the outer boundary of the domain \citep{Mih78}.
However, to reproduce the analytic solution for an infinitely extended
domain, we can take the layers beyond the computational domain into
account if the flux is assumed to be known and the solution is in radiative
equilibrium.
This is explained in appendix~\ref{ImprovedOuterBC}.

The code is parallelised by splitting the problem into two local ones
and a nonlocal one in between.
The latter requires interprocessor communication.
The local problems are computationally intensive, while the nonlocal 
problem does not involve any computations and is therefore quite fast.
On each processor, one first solves \eq{ngradI_eqn} along each ray to 
compute the intrinsic intensity increment within each processor as a 
function of optical depth.
In the second step, the increments of intrinsic intensity and optical
depth are communicated to the neighbouring processors.
In the third and final step, these increments are used to construct the
total intensity within each processor.
For most of the calculations presented here, we use just the vertical
or $z$ direction, which corresponds to two ray directions, namely for upward 
and downward propagating radiation.

The radiative time step problem was never addressed in this code,
because in the applications of \cite{HNSS07} to solar convection and
sunspot formation, no severe time step constraints were encountered.
Doing this quantitatively is the main purpose of the present paper.

\subsection{Radiative time steps from an unstratified model}
\label{Unstratified}

To have a preliminary idea about the values of
$C_{\rm rad}^{\rm thick}$ and $C_{\rm rad}^{\rm thin}$,
we perform simple one-dimensional experiments by solving
\Eqs{ngradI_eqn}{opt_thickthin} for constant $\rho$ and
$\kappa$ (here with $\ell\equiv(\kappa\rho)^{-1}=1\Mm$) for
an unstratified model.
We adopt periodic boundary conditions for $I$ and $T$ and look for
the decay of a sinusoidal temperature perturbation of the form
\EQ
T(z,t=0)\,=\,T_0+T_1\sin kz,
\EN
where $T_0=10^6\K$ is chosen hot enough so that radiation provides the
most restrictive time step constraint and $T_1=10^5\K$.
The density is chosen to be $10^{-9}\g\cm^{-3}$.
The value of $c_\gamma$ is then $2.2\times10^{10}\km\s^{-1}$.
This exceeds the speed of light, but that does not matter in this
non-relativistic computation, and it helps making sure the acoustic
time step remains unimportant under all circumstances.
Thus, in the optically thin case, the cooling time $\ell/c_\gamma$
is much smaller than the sound wave crossing time, $\delta z/\cs$.

Here we decided to retain \Eqs{dlnrho}{DuDt}, which
implies that sound waves can equilibrate pressure perturbations
on an acoustic time scale.
This has another interesting side effect in that the
temperature perturbations then tend to cause corresponding
density variations rather than pressure variations.
This justifies in hindsight that the $\DD p/\DD t$ term in
\eq{opt_thickthin} could then be omitted.
This, in turn, justifies the presence of $\cp$ instead of
$\cv$ in the definition of $c_\gamma$ in \eq{cgam_def};
see appendix~\ref{SpiegelVeronis} for details.

\begin{figure*}[t!]\begin{center}
\includegraphics[width=\textwidth]{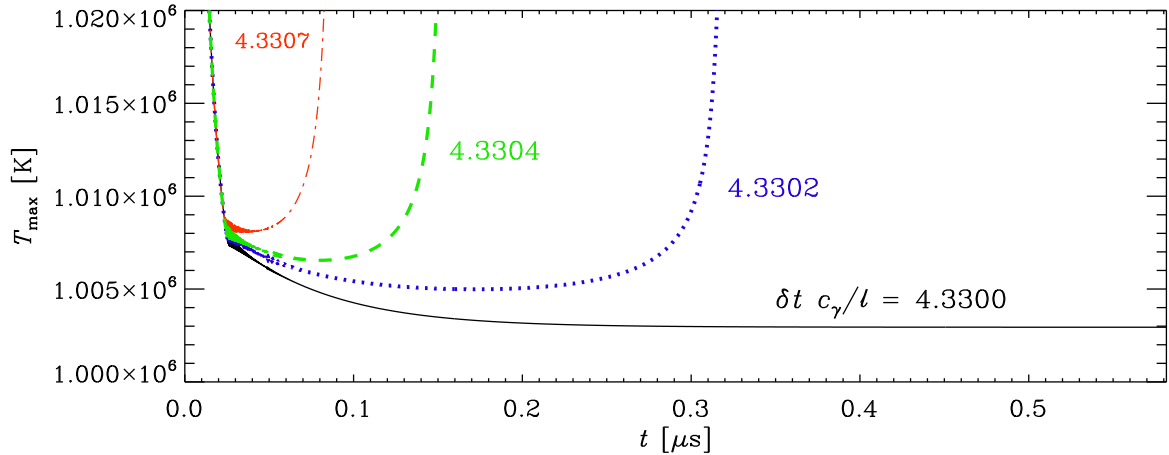}
\end{center}\caption[]{
$T_{\max}(t)$ for different values of
$\delta t\,c_\gamma/\ell$, for the unstratified model of \Sec{Unstratified}
(colour online).}\label{pdt_blow}\end{figure*}

We vary $k$ in the range from $10^{-4}$ to $10\Mm^{-1}$ and
choose $N=128$ mesh points, so the domain mesh widths are
$\delta z=2\pi/kN$, which varies
between approximately $500$ and $0.005\Mm$, respectively.
In all cases, we determine the maximum permissible time step $\delta t$.
An example of the time evolution of the maximum temperature $T_{\max}$
versus time is presented in \Fig{pdt_blow}, which show that
$T_{\max}$ develops runaway when the time step is too long.
In this case, we find that $C^{\rm thin}_{\rm rad}$ or  
$\delta t\,c_\gamma/\ell\approx4.3300$
is the approximate borderline. Already 4.3302 is too large.

\begin{figure*}[t!]\begin{center}
\includegraphics[width=\textwidth]{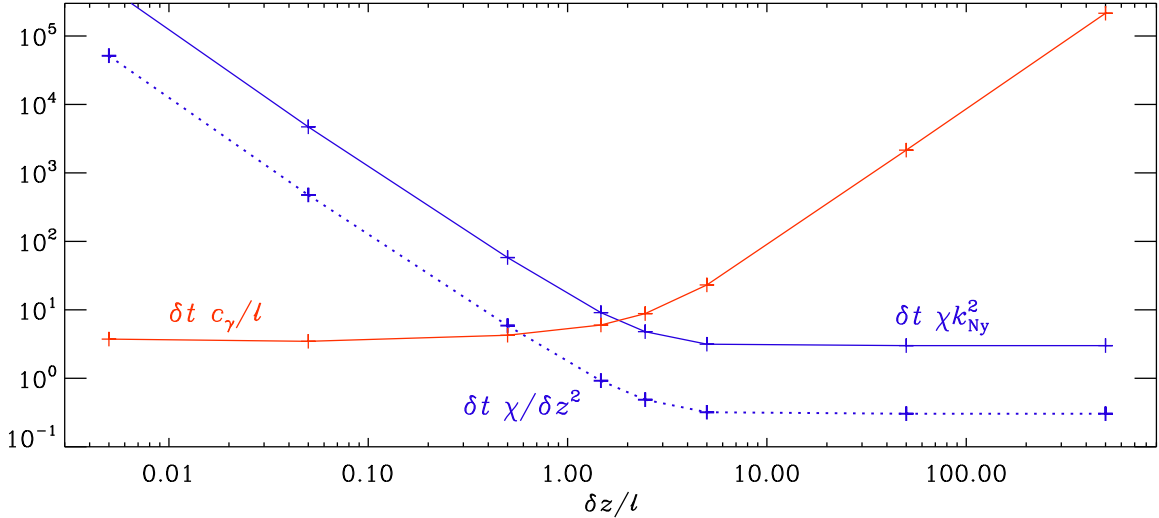}
\end{center}\caption[]{
The maximum permissible time step $\delta t$ versus $\delta z/\ell$,
normalised by $\ell/c_\gamma$ (red) and by $(\chi k_{\rm Ny}^2)^{-1}$
(solid blue line), as well as $\delta z^2/\chi$ (dotted blue line),
for the unstratified one-dimensional model of \Sec{Unstratified}.
Note that $\delta t\,\chi k_{\rm Ny}^2=\delta t\,\chi\pi^2/\delta z^2$
(colour online).}\label{pdt_table}\end{figure*}

In \Fig{pdt_table}, we show the resulting maximum permissible time step,
$\delta t$, in two different normalisations related to the two terms
of \Eq{taucool}.
When $\delta t$ is normalised by $\ell/c_\gamma$, which is a constant in
this model, we see that $\delta t$ increases quadratically with $\delta z$
for large values, but is independent of $\delta z$ for small values -- 
saturating at a value of about $3$.
This is in perfect agreement with \Eq{dtrad_new}.
On the other hand, when $\delta t$ is normalised by $\delta z^2/\chi$,
it levels off at a value of around 0.3.
Normalising instead by $(\chi k_{\rm Ny}^2)^{-1}$, we see that the
maximally permissible time step levels off at a value
of about $3$, i.e., same as that for the normalisation by $\ell/c_\gamma$.
This may justify the use of a single coefficient for the radiation time step,
as done by \cite{FSL12}.
However, since the values of the two coefficients depend on the numerical
scheme, we retain the two independently.

For small $\delta z$ (optically thin limit), we see that 
$\delta t\,c_\gamma/\ell\approx3$--$4$, while for large $\delta z$
(optically thick limit), we obtain $\delta t\,\chi/\delta z^2\approx0.3$.
These, then, would be the recommended values of $C_{\rm rad}^{\rm thin}$
and $C_{\rm rad}^{\rm thick}$, respectively.
However, throughout this work, we adopt a more conservative value for
the latter, $C_{\rm rad}^{\rm thick}=0.2$, which is later found to be
necessary for numerical stability in some stratified cases, while
$C_{\rm rad}^{\rm thin}\approx4$ turns out to work well in the cases
presented below.

\begin{table}[b!]\caption{
Values of $\delta t\,\chi/\delta z^2$ for the shortest permissible time
step for given values of the number of dimensions $D$ and the number of
rays $n_{\rm ray}$ in the optically thick regime.
}\vspace{12pt}\centerline{\begin{tabular}{c|ccccc}
\hline
\hline
$D$ & 1 & 2 & 3 & 3 & 3 \\
\hline
$n_{\rm ray}$ & 2 & 4 & 6 & 14 & 22 \\
\hline
$\delta t\,\chi/\delta z^2$ &
$0.375\pm0.001$ &
$0.188\pm0.001$ & 
$0.127\pm0.005$ & 
$0.218\pm0.005$ & 
$0.291\pm0.005$ \\ 
\label{Tdir}\end{tabular}}\end{table}

Let us now comment on the spatial properties of the solution in the
optically thick and thin cases when the numerical instability develops,
i.e., when the time step is too long.
For $\delta z/\ell\gg1$, i.e., when the time step is constrained by
$\delta z^2/\chi$, the numerical instability
develops its fastest growing mode uniformly over the domain
at the mesh scale, i.e., at $k=k_{\rm Ny}$.
This does indeed correspond to a checkerboard pattern in two dimensions,
as was assumed in our analysis above.
For $\delta z/\ell\ll1$, on the other hand, the fastest growing mode
also develops at $k=k_{\rm Ny}$, but nonuniformly and preferentially
in those locations that are hotter.

We have stated in \Sec{RadiativeCoolingConstraint} that the time step
in the optically thick regime scales with the dimension $D$.
This is demonstrated in \Tab{Tdir}, where we compare the maximum permissible
$\delta t$ in units of $\delta z^2/\chi$ in multidimensional domains
($D=1$, $2$, and $3$) for different numbers of rays, $n_{\rm ray}$.
In those cases, the mesh spacing is the same in all directions.
We see that the time step is approximately inversely proportional to $D$
if we restrict ourselves to rays along the coordinate direction, i.e.,
$n_{\rm ray}=2$, $4$, or $6$.
In that case, $\delta t\,\chi/\delta z^2\approx0.38/D$.
Interestingly, when more directions are included ($n_{\rm ray}=14$ or $22$),
the minimal timestep becomes longer again.
This is because the radiative flux divergence is calculated
as an angular integral over all directions.
However, the diagonal directions do not contribute for a checkerboard
pattern.
Therefore, the radiative flux divergence decreases for larger $n_{\rm ray}$,
which alleviates the time step constraint correspondingly.
It turns out that the extra cost associated with the use of more rays
is easily being outweighted by being able to use a longer time step.
The increased accuracy obtained by using more rays comes therefore
effectively at no extra cost.

\subsection{Stratification of hot stellar surface layers}
\label{MLTmodel}

Later in this paper, we address the question which time step
constraint plays a role in which layers of certain stars.
For that purpose, we need a simple model for stellar surface layers.
We therefore present here the relevant equations that can simply be
solved by integration.
The same solutions can also be obtained using explicit time integration
with the {\sc Pencil Code}.

In hot stellar surface layers, the radiation pressure plays an
important role.
This is typically also the regime in which the electron scattering
opacity is important \citep{FKR92}, i.e.,
\EQ
\kappa=\kappa_{\rm es}\approx 0.34\cm^2\g^{-1}.
\label{kappa_es}
\EN
Since this is a constant, and since the radiative flux $F_{\rm rad}$ is
also constant in radiative equilibrium, we just have to replace gravity
by the effective one, which is then still a positive constant, i.e.,
\EQ
g\to g_{\rm eff}=g-(\kappa/c)\,F_{\rm rad}=\const.
\label{eq_geff}
\EN
To obtain a hydrostatic reference solution, we integrate the
equations of hydrostatic and radiative equilibrium,
\EQ
{\dd p\over\dd z}=-\rho \, g_{\rm eff},\qquad
{\dd T\over\dd z}=-{F_{\rm rad}\over K},
\label{dpdT_eqns}
\EN
with $F_{\rm rad}=\sigmaSB\Teff^4=\const$, where $\Teff$ is the effective 
temperature of the stellar surface and $K=\rho \cp \chi$.
Note that the equation for $\dd T/\dd z$ applies even in the optically
thin case provided the system is in thermal equilibrium.
This is because in equilibrium we have $\nab \bm \cdot\FF_{\rm rad}=0$,
and therefore $J=S$, just like in the optically thick case.
In the time-dependent case, however, as discussed above, the optically
thick and thin cases are quite different from each other.
It is only in that case that a time step constraint has to be obeyed.

Next, it is convenient to divide the two equations in \eq{dpdT_eqns}
by each other, so we obtain,
\EQ
{\dd T\over\dd p}={F_{\rm rad}\over K\rho \, g_{\rm eff}}.
\EN
In astrophysics, the symbol $\nabla$ is used to denote the double-logarithmic
temperature gradient, so $\nabla=\dd\ln T/\dd\ln p$, and for the present
radiative equilibrium solution, the radiative temperature gradient
is denoted by $\nabla_{\rm rad}$.
Thus, we have
\EQ
\nabla_{\rm rad}\equiv{\dd\ln T\over\dd\ln p}
={p F_{\rm rad}\over KT\rho \, g_{\rm eff}}
={(\cp-\cv)\,F_{\rm rad}\over K \, g_{\rm eff}},
\label{nabla_rad}
\EN
where we have used $p=(\cp-\cv)\,T\rho$ for the equation of state
of a perfect gas and $\cv$ is the specific heat at constant volume.

The vertical specific entropy gradient is
$\dd(s/\cp)/\dd\ln p=\nabla-\nabla_{\rm ad}$, so the Schwarzschild
criterion for convective stability (positive outward gradient of $s$)
is $\nabla<\nabla_{\rm ad}$.
Convection occurs when $\nabla>\nabla_{\rm ad}$.
This mixes the fluid, so $s$ becomes uniform and one must
replace $\nabla$ by $\nabla_{\rm ad}=1-1/\gamma=0.4$, which is
the value for a monatomic gas with $\gamma=\cp/\cv=5/3$.
Thus, the local double-logarithmic temperature gradient can be written as
\citep[see, e.g.,][]{Kah72}
\EQ
\nabla=\min(\nabla_{\rm ad},\nabla_{\rm rad}).
\label{nabla_min}
\EN
In more realistic mixing length descriptions of stellar convection, this
relation is to be replaced by a smooth transition between the two
states; see \cite{Vit53} for the original formulation, which corresponds
to finding a solution to the equation
\EQ
(\nabla-\nabrad)+\epsilon_\ast(\nabla-\nabrad)^\xi=0
\qquad\mbox{(for $\nabrad>\nabad$)}
\label{nabla_eqn}
\EN
with $\xi=3/2$, $\epsilon_\ast=\iota\cs^3/\chi g$, and $\iota$ being
a coefficient of the order of unity; see figure~3 of \cite{Bra16} for
a comparison of solutions for $\xi=3/2$ and $\xi=1$.

For lower temperatures, the opacity given by \eq{kappa_es}
is no longer valid.
A more general representation is given in terms of combinations
of Kramers-type opacities
\EQ
{1\over\kappa}={1\over\kappa_{\rm H^-}}+{1\over\kappa_{\rm Kr}+\kappa_{\rm es}},
\label{kappa_tot}
\EN
where we use
\EQ
\kappa_i=\kappa_0 \, (\rho/\rho_0)^{a_i} (T/T_0)^{b_i}
\label{Kramers}
\EN
with $i={\rm Kr}$ or ${\rm H^-}$. For our hot stellar surface models, we use
$\kappa_0=10^4\cm^2\g^{-1}$, $\rho_0=10^{-5}\g\cm^{-3}$,
$T_0=13,000\K$, and $a_{\rm Kr}=1$, $b_{\rm Kr}=-3.5$ for the Kramers
opacity $\kappa_{\rm Kr}$, relevant for the deeper layers in the star, 
and $a_{\rm H^-}=0.5$, $b_{\rm H^-}=18$ for the 
H$^-$ opacity $\kappa_{\rm H^-}$ in the layers just beneath the photosphere.
These were also the coefficients used by \cite{Bra16}.
The same opacity prescription will also be used in our accretion disc
models described below, but with $\kappa_0=2\times 10^4\cm^2\g^{-1}$, 
$T_0=20,000\K$, $a_{\rm H^-}=1$ and $b_{\rm H^-}=4$.

To construct a solution, we assume $\Teff$ and $g_{\rm eff}$ as given.
For grey atmospheres, the temperature of the atmosphere far above the
photosphere is $T_0=\Teff/2^{1/4}\approx0.84\,\Teff$; see \cite{Sti02}
for a text book.
Thus, we integrate from the top downward using $\ln p$ as the independent
variable starting with a sufficiently low value.
For single power law opacities, such as a simple Kramers-type opacity with
fixed exponents $a$ and $b$, this integration can be done analytically
\citep[see appendix~A of][]{Bra16}, but here we use more complicated
opacities and do the integration numerically.
At each height, we solve \Eq{nabla_eqn} for $\nabla$ to determine the
temperature gradient when $\nabrad>\nabad$, in which case we replace
$\nabla$ by  $\nabad$.

\subsection{Accretion disc model}
\label{AccretionDiscModel}

Accretion discs are generally heated by turbulent dissipation.
To model this realistically, we would need to simulate accretion disc
turbulence through a combination of (i) the magneto-rotational instability
(MRI) to generate turbulence from a given magnetic field \citep{BH91}
and (ii) the dynamo instability to regenerate the required magnetic
field \citep{BNST95}.
To simplify matters, and to avoid modelling the MRI and the dynamo,
we assume instead a prescribed heating function as a function of $z$,
\begin{equation}
{\cal H}(z)=\left(\threehalf\varOmega\right)^2 {\dot{M}\over3\pi} \,
{\varTheta(z_{\rm heat}-|z|)\over z_{\rm heat}},
\label{eq_heatfunc}
\end{equation}
where we have used the usual parameterisation for accretion discs in
terms of mass accretion rate $\dot{M}$ and local Keplerian angular
velocity $\varOmega$ \citep{FKR92}.
In addition, we have assumed a vertical profile $\varTheta(z_{\rm heat}-|z|)$,
which is unity inside the disc, $|z|\leq z_{\rm heat}$, and zero outside.
In practice, we initiate our simulations by using an isothermal
hydrostatic stratification, where $g=\varOmega^2z$ is the vertical gravity
and $\ln(\rho/\rho_0)=-z^2/2\Hp^2$, with $\Hp=\cs/\varOmega$ being the
pressure scale height, $\cs=\sqrt{\Rgas T/\mu}$ the isothermal sound
speed, $\Rgas$ the universal gas constant, $\rho_0$ 
the density at $z=0$, and $\mu$ the mean atomic weight.
Radiation then causes the outer layers to cool until thermal 
equilibrium is achieved.
Alternatively, one could start with a thermal hydrostatic equilibrium
that is computed using \Eqss{dpdT_eqns}{nabla_rad}, except that now
$F_{\rm rad}\neq\const$. Hence, this has to be obtained by integrating
$\dd F_{\rm rad}/\dd z={\cal H}(z)$ with $F_{\rm rad}(z=0)=0$ as
a boundary condition at the midplane.

For the purpose of understanding the radiative time step constraint,
we will not be concerned with convection in these simulations with
the {\sc Pencil Code}.
Instead, we use a one-dimensional model in which the flow is either up
or down, but not both, so no return flow and no convection are possible.
Convection will, however, be discussed briefly in \Sec{3DdiscDNS},
where we illuminate in more detail the properties of discs around
white dwarfs.
In such discs, the total flux is given by 
$F_{\rm tot} = F_{\rm rad} + F_{\rm conv}$, and it is 
found that a significant fraction of the convective flux $F_{\rm conv}$ is
independent of the superadiabatic gradient, as is normally assumed
in standard mixing length theory.
This is particularly interesting in view of recent suggestions
\citep{Bra16} that even in solar and stellar convection, the convective
flux may have a significant contribution from what is known as the
\cite{Dea66,Dea72} term.
Numerical evidence for this flux was found in simulations of stellar
convection \citep{Kapy_etal17,KVKB18}, and the convective simulations 
of accretion discs in this work will provide further evidence for this.

Note that we have not made any attempt to smoothen the abrupt change in
the heating function of \eq{eq_heatfunc}
and have rather regarded this feature as an advantage. This is because
it allows us to see whether this profile results in a similarly abrupt 
transition in the resulting vertical profiles of temperature or radiative flux.
As will be demonstrated in \Sec{DNSdiscmodels}, the temperature
profile turns out to be smooth, suggesting that no artefacts of the
heating profile are introduced into the model.
The total flux $F_{\rm tot}$, on the other hand, shows a sharp first derivative
in the steady state. Again, this is not a problem, since it serves 
as a convenient ``marker'' of where the heating stops in the vertical 
profiles of various fluxes. We will return to this in \Sec{3DdiscDNS}. 

Both in the stellar and in the accretion disc models, we determine
the location of the photosphere as the point where the optical depth,
\EQ
\tau(z)=\int_z^\infty\kappa(z')\rho(z')\,\dd z',
\EN
is unity.
Here, $z\to\infty$ corresponds to a location far away from the disc
or the star, but for the lower disc plane, the integral would need
to go from $-\infty$ to $z$ instead.

\section{Results}

\subsection{A hot stellar surface layer}
\label{sec_hotstar}

\begin{figure*}[t!]\begin{center}
\includegraphics[width=\textwidth]{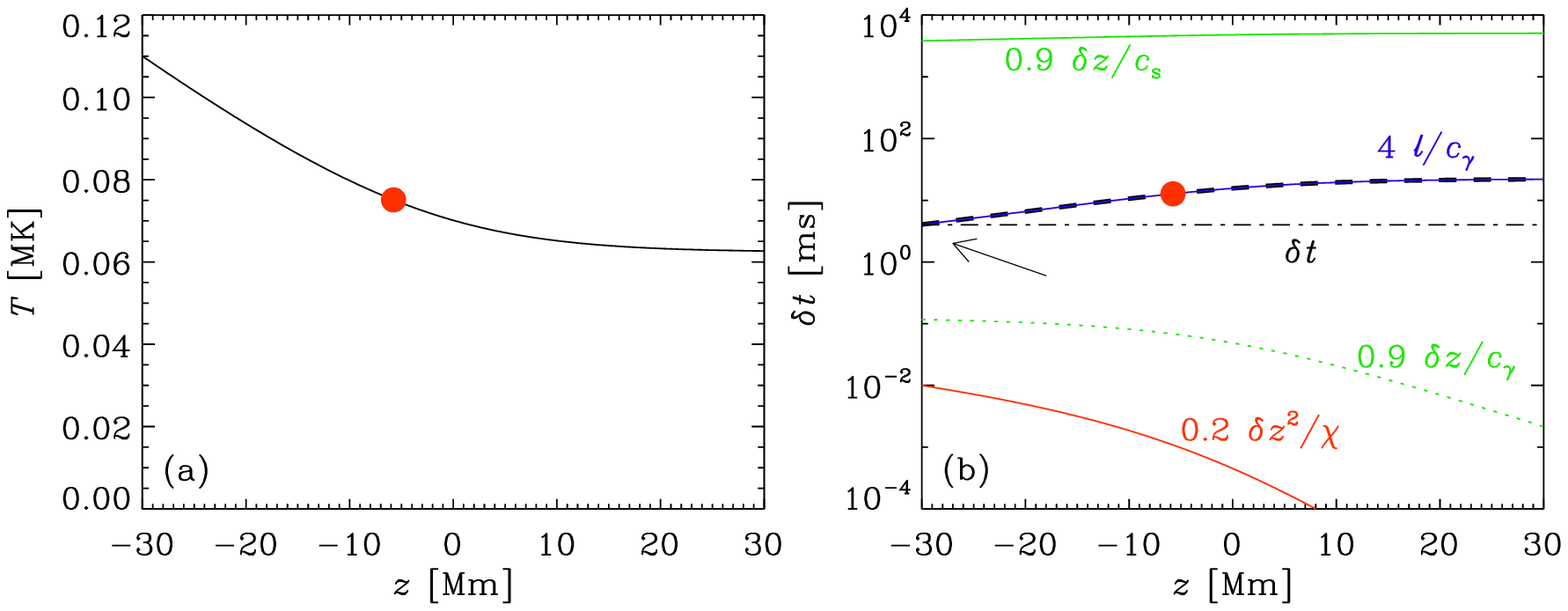}
\end{center}\caption[]{
(a) Temperature stratification from the {\sc Pencil Code} simulation of
a hot star with $T_{\rm eff}=69000K$ (see \Sec{sec_hotstar}).
(b) $z$ dependence of various time step constraints: 
$\delta t_{\rm rad}^{\rm thick}$ (red solid line), 
$\delta t_{\rm rad}^{\rm thin}$ (blue solid line), 
$\delta t_{\rm rad}$ (black dashed line), $\delta t_{\rm s}$
(green solid line), $\delta t_\gamma$ (green dotted line), and 
the empirically determined maximum permissible time step $\delta t$ 
(black dot-dashed line). The red dot denotes the photosphere.
The arrow points to the location where the minimum of $4\,\ell/c_\gamma$
coincides with $\delta t$ and is therefore constraining the time step.
All time steps are in milliseconds
(colour online).}\label{pdt_L256b2_axel1}\end{figure*}

We adopt here one of the stellar surface layer models of
the unpublished work of Brandenburg and Spiegel, who considered
a star of solar mass $M=M_\odot=2\times10^{33}\g$,
solar radius $R=R_\odot=7\times10^{10}\cm$ and 
$g=g_\odot=GM_\odot/R_\odot^2=2.7\times10^4\cm\s^{-2}$ (the solar value),
but with a luminosity $L$ that is $2\times10^4$ times the solar value 
$L_\odot=4\times10^{33}\erg\s^{-1}$.
The radiative flux is given by $F_{\rm rad}=L/(4\pi R^2)
=1.3\times10^{15}\erg\cm^{-2}\s^{-1}$ and the
effective temperature $\Teff=69,000\K$.
We solve the time-dependent \Eqss{dlnrho}{ngradI} for 
the surface layers of the above star using the {\sc Pencil Code}, 
with initial $\uu=0$. We also put ${\cal H}=0$,
$\nu=10^{12}\cm^{2}\s^{-1}$, and $C_{\rm shock}=3$.
We refer to appendix~\ref{ImprovedOuterBC} regarding the boundary
conditions imposed on the intensity, and appendix~\ref{hydrobc} for those  
on the hydrodynamic variables.
Our model has a depth of $60\Mm$ and uses 256 uniformly spaced mesh points.
The $\tau=1$ surface
is roughly in the middle of the domain, which we define to be at $z=0$.
The temperature then varies between $110,000\K$ at $z=-30\Mm$
and $62,000\K$ at $z=30\Mm$.
The sound speed varies between $\cs=55\km\s^{-1}$ at the bottom
and $\cs=42\km\s^{-1}$ at the top, while $c_\gamma=55\km\s^{-1}$
at the bottom and about $10^8\km\s^{-1}$ at the top.
The domain has a density contrast of $\rho_{\max}/\rho_{\min}\approx300$,
so the number of density scale heights is $\Delta\ln\rho=\ln300\approx6$.
Note that, for this case, we have used $\kappa=\kappa_{\rm es}=\const$ 
(i.e., in \Fig{pdt_L256b2_axel1} only) instead
of expression \eq{kappa_tot}, but the difference would
be minor.

In \Fig{pdt_L256b2_axel1}, we show the steady state temperature
stratification for the above star (left panel), and the various candidates
that could play a role in constraining the time step (right panel).
The total duration of this simulation was $56.4 \s$. 
Note that this figure, as well as the rest of the figures obtained 
from {\sc Pencil Code} simulations in this paper, 
have been plotted at the final time of the corresponding 
simulation (by which the solution has 
attained a steady state), unless otherwise mentioned.
We also compare with the empirically determined maximum permissible 
time step $\delta t$ obtained in the {\sc Pencil Code} 
simulations, which turns out to be $4 \ms$ for this case.
Note that every simulation is run with a constant, pre-assigned time step. 
We recall that, in any time-dependent calculation, regardless of whether
or not a steady state solution is reached, the simulation may become
numerically unstable if the time step is too long.
One can then keep decreasing the supplied time step (to the desired 
accuracy) in order to determine the limiting time step for that particular 
problem, namely $\delta t$ in our notation.

Interestingly, we find in \Fig{pdt_L256b2_axel1} that 
$\delta t$ is very close to the line for 
$\delta t_{\rm rad}^{\rm thin}=4\,\ell/c_\gamma$.
The line for $\delta t_{\rm rad}^{\rm thick} = 0.2\,\delta z^2/\chi$ is 
well below the actual $\delta t$, which is consistent with \eq{dtfull}
in that a term proportional to $\delta t_{\rm rad}^{\rm thick}$ does 
not constrain $\delta t$ in isolation.
Similarly, $\delta t_\gamma=0.9\,\delta z/c_\gamma$ is not
motivated to be a possible candidate for constraining the time step 
in this case.
The acoustic time step $\delta t_{\rm s}=0.9\,\delta z/\cs$ is much larger
than the radiative one and therefore unimportant in this case.
It is important to note here that the right-hand side of \eq{dtfull_wrong}
would have predicted $\delta t_{\rm rad}^{\rm thick}$ to be the 
relevant time step in the problem. However, in reality it turns out 
to be $\delta t_{\rm rad}^{\rm thin}$, as also supported 
by our numerical simulations. 

\cite{ST99} motivated the interest in studying the hydrodynamics of hot
stars by referring to O.\ Struve for having discovered large line widths,
which, in turn, could hint at the existence of turbulence in the
atmospheres of those stars.
They associated this line broadening with photofluid instabilities,
which are possible even when the radiative acceleration is still below
the gravitational one.
In the present case, the radiative acceleration is $0.53$ times the
gravitational one, and may even exceed it at higher temperatures.
This could lead to photoconvection \citep{PS73,Spi77,ST99}, which
Brandenburg and Spiegel attempted to study with the
{\sc Pencil Code}, but suspended this project because of prohibitively
short time steps.
In recent years, similar studies have been performed in the context
of radiation-driven stellar winds \citep{OS18,SOP18}.
No time step problems, have been reported in these studies.
However, we point out that the radiation is 
treated differently there, as they consider two-dimensional radiation 
line-transport. Also, 
the time step is chosen such that it is 
the minimum of a fixed time and a variable 1/3 of the Courant time.

\subsection{Expected time step constraints for stellar surface layers}
\label{StellarSurfaceLayers}

\begin{figure*}[t!]\begin{center}
\includegraphics[width=\textwidth]{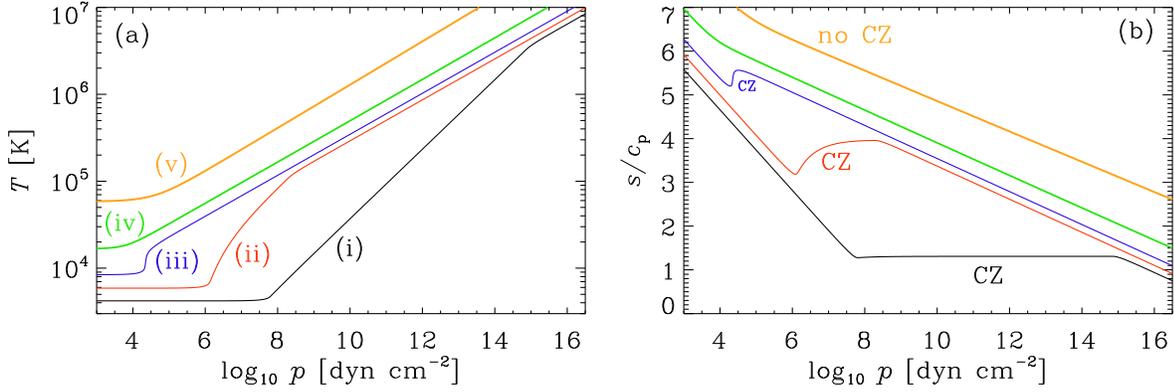}
\end{center}\caption[]{
Temperature (a) and specific entropy (b) stratification 
from the mixing length model discussed in \Sec{MLTmodel}
for $\Teff=5000\K$ (i), $7000\K$ (ii), $10,000\K$ (iii), $20,000\K$ (iv) 
and $\Teff=70000\K$ (v),  using $g=2.7\times10^4\cm\s^{-2}$.
The locations of the convection zones are marked by CZ in three of the
five curves where $\dd s/\dd\ln p >0$, corresponding to a positive
superadiabatic gradient
(colour online).}\label{pdt_2panels}\end{figure*}

\begin{figure*}[t!]\begin{center}
\includegraphics[width=\textwidth]{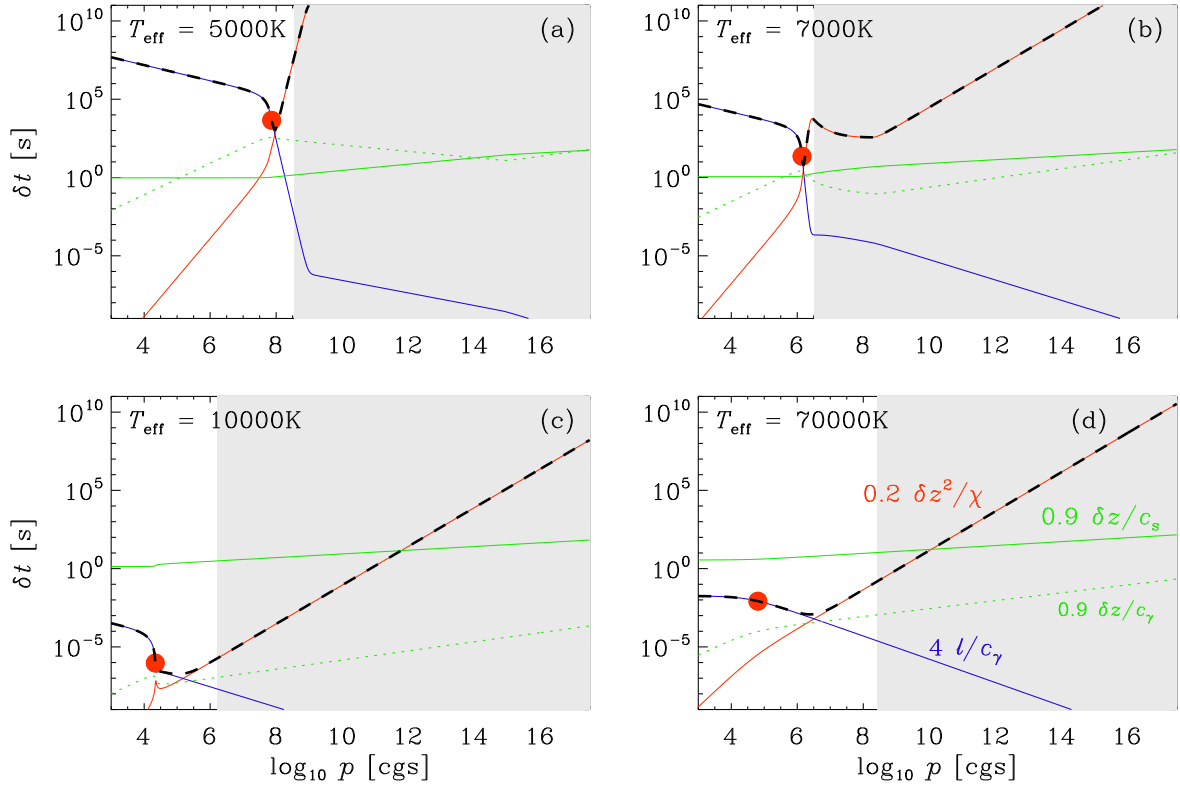}
\end{center}\caption[]{
Time step constraint for the mixing length models (i) in panel (a),
(ii) in panel (b), (iii) in panel (c) and (v) in panel (d):
$\delta t_{\rm rad}$ (black dashed lines),
$\delta t_{\rm rad}^{\rm thin}$ (blue solid lines),
$\delta t_{\rm rad}^{\rm thick}$ (red solid lines), 
$\delta t_{\rm s}$ (green solid lines), and
$\delta t_\gamma$ (green dotted lines). 
The grey regions denote the regime of dynamic diffusion 
in stars, i.e., where $\tau \gg 1$ and $\beta \tau \gg 1$.
All time steps are in seconds
(colour online).}\label{pdt_4panels}\end{figure*}

We now consider solutions of the time-independent \Eqss{eq_geff}{kappa_tot}
for stellar surface layers (see \Sec{MLTmodel} for the method).
So no time step constraint applies, but we can still use this model
to predict what the time step constraint would be in a time-dependent
simulation.
In \Fig{pdt_2panels}, we plot $T(\ln p)$ and $s(\ln p)$ for five values
of $\Teff = 5000\K, 7000\K, 10000\K, 20000\K$ and $70000\K$.
We use $g=g_\odot$ for the following plots.
In \Fig{pdt_4panels}, we plot the two contributions to the radiative
time step, as well as the acoustic one,
\EQ
\delta t_{\rm rad}^{\rm thick}=C_{\rm rad}^{\rm thick}\,{\delta z^2\over\chi},
\qquad
\delta t_{\rm rad}^{\rm thin}=C_{\rm rad}^{\rm thin}{\ell\over c_\gamma},
\qquad
\delta t_{\rm s} = C_{\rm CFL} \frac{\delta z}{\cs},
\EN
together with the total radiative one $\delta t_{\rm rad}=
\delta t_{\rm rad}^{\rm thick} + \delta t_{\rm rad}^{\rm thin}$ for 
$\delta z=0.05\Hp$, where $\Hp = (c_{\rm p} - c_{\rm v})T/g$ 
is the local pressure scale-height in the star.
In all of our models, the temperature above the photosphere reaches
a constant; see \Fig{pdt_2panels}(a).
This is because the physics of realistic coronal heating and cooling is
not included in our simple model; see the papers by \cite{BP11} and
\cite{BBP13}
for realistic coronal modelling with the {\sc Pencil Code}, using a setup
originally developed by \cite{GN02,GN05a,GN05b}.

Next, we discuss the constraints on the resulting time step,
that would limit a time-dependent calculation of the same problem.
In \Fig{pdt_4panels}, at small values of $\ln p$,
corresponding to locations above the photosphere, we have
$\delta t_{\rm rad}^{\rm thin}\gg\delta t_{\rm rad}^{\rm thick}$,
so their sum is determined by $\delta t_{\rm rad}^{\rm thin}$, as expected;
see \Fig{pdt_4panels} for small values of $\ln p$, where the blue line for
$\delta t_{\rm rad}^{\rm thin}$ is the highest.
In the deeper layers below the photosphere, the situation is the
other way around and the time step is expected to be governed by
$\delta t_{\rm rad}^{\rm thick}$, again as expected; see \Fig{pdt_4panels}
for large values of $\ln p$ (i.e., below the photosphere), where the
red line for $\delta t_{\rm rad}^{\rm thick}$ is the highest.
The radiative time step is thus determined by the minimum of
$\delta t_{\rm rad}$, corresponding to the black dashed line in
\Fig{pdt_4panels}. 
We find that $\delta t_{\rm rad}$ is the shortest either in the photosphere
[panels (a) and (b) of \Fig{pdt_4panels}],
indicated by a red dot where $\tau=1$, or just below it [see panels (c) and (d)
of \Fig{pdt_4panels}, i.e., for models with shallow or no outer convection zone].

Note that our restriction
of static diffusion may no longer be justified if $\beta \tau \gg 1$ 
in the deeper layer of the stars, 
which are known to be in the dynamic diffusion regime; see \cite{Kru07}. 
To estimate the depth where this occurs in the stars in \Fig{pdt_4panels}, 
let us estimate $u$ based on the
assumption that the convective flux can be modelled by the associated
mixing length expression, $\sigmaSB\Teff^4=\rho u^3$.
The depth below which dynamic diffusion occurs is obtained when
$\beta \tau \approx(\sigmaSB\Teff^4/\rho)^{1/3}\tau/c$ exceeds unity.
In that regime, our radiation transport equations are no longer
applicable, which is denoted by the grey regions in \Fig{pdt_4panels}.
We see that $\beta\tau\ll1$ is always obeyed in the surface layers of
all the stars, where our radiation transport equations are valid.
Note that our estimate of $\beta\tau$ applies only to regions where
convection is possible, i.e., in the outer layers of cool stars or in the
convective cores of hot stars.
Thus, the actual depth where $\beta \tau \approx 1$ may differ.

\begin{figure*}[t!]\begin{center}
\includegraphics[width=\textwidth]{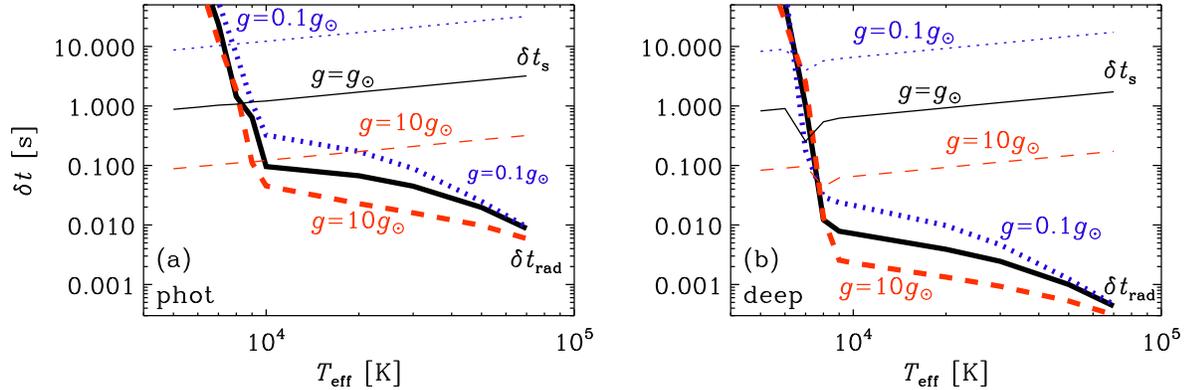}
\end{center}\caption[]{
$\delta t_{\rm rad}$ and $\delta t_{\rm s}$ at the photosphere
(a) and at the location where $\delta t_{\rm rad}$ is minimum (b)
for $g=g_\odot=2.7\times10^4\cm\s^{-2}$ (black solid lines),
$g=0.1\,g_\odot$ (blue dotted lines), and $g=10\,g_\odot$ (red dashed lines).
The thick and thin lines correspond to $\delta t_{\rm rad}$ 
and $\delta t_{\rm s}$ respectively.
All time steps are in seconds
(colour online).}\label{presults}\end{figure*}

The maximum permissible time step $\delta t$ for this problem is
then given by the minimum of $\delta t_{\rm rad}$
and the acoustic time step $\delta t_{\rm s}$ (the green solid line).
As can be seen from \Fig{pdt_4panels}, for $\Teff=5000\K$, it is
always $\delta t_{\rm s}$ that constrains the actual time step.
Although for $\Teff=7000\K$ the minimum of $\delta t_{\rm rad}$
occurs below the photosphere, at that location 
$\delta t_{\rm s}$ is still the shorter time step.
However, for $\Teff=10,000\K$ and higher, $\delta t_{\rm s}$ is no
longer the constraining time step, instead it is $\min(\delta t_{\rm rad})$.
To better understand the dependence of various time steps 
on $\Teff$ and $g$, we turn to \Tab{THR} and \Fig{presults}.
In \Tab{THR}, we give the radiative and acoustic time steps
in the photosphere, $\delta t_{\rm rad}^{\rm phot}$ and
$\delta t_{\rm s}^{\rm phot}$, respectively, and the radiative
and acoustic time steps at the position where the radiative time step
has a minimum, $\delta t_{\rm rad}^{\rm deep}$ and
$\delta t_{\rm s}^{\rm deep}$, respectively; for the five
models (i)--(v) discussed above in \Figs{pdt_2panels}{pdt_4panels},
and for a few other ones.
In \Fig{presults}, we plot these newly defined time steps as a function of 
$\Teff$ for different values of $g$. 

\begin{table}[t!]\caption{
$\delta t_{\rm rad}$ and $\delta t_{\rm s}$ at the photosphere
(superscript `phot') and at the location where $\delta t_{\rm rad}$ is
minimum (superscript `deep') for some values of $\Teff$ and $g$.
All time steps are given in seconds.
}\vspace{12pt}\centerline{\begin{tabular}{crc|cc|cc}
Model & $\Teff\;$  & $g/g_\odot$ &
$\delta t_{\rm rad}^{\rm phot}$ &  $\delta t_{\rm s}^{\rm phot}$ &
$\delta t_{\rm rad}^{\rm deep}$ &  $\delta t_{\rm s}^{\rm deep}$ \\
\hline
 (i) & 5,000 & 1 & $4.7\times10^{+3}$ & 0.88 & $8.5\times10^{+2}$ & 0.83 \\
     & 6,000 & 1 & $2.5\times10^{+2}$ & 0.95 & $5.4\times10^{+1}$ & 0.90 \\
     & 7,000 & 1 & $2.4\times10^{+1}$ & 1.03 & $1.0\times10^{+0}$ & 0.25 \\
(ii) & 8,000 & 1 & $1.4\times10^{+0}$ & 1.07 & $1.2\times10^{-2}$ & 0.54 \\
     & 9,000 & 1 & $6.4\times10^{-1}$ & 1.17 & $7.9\times10^{-3}$ & 0.62 \\
(iii)&10,000 & 1 & $9.6\times10^{-2}$ & 1.21 & $7.2\times10^{-3}$ & 0.66 \\
(iv) &20,000 & 1 & $6.7\times10^{-2}$ & 1.71 & $3.9\times10^{-3}$ & 0.93 \\
     &30,000 & 1 & $4.5\times10^{-2}$ & 2.09 & $2.5\times10^{-3}$ & 1.14 \\
     &50,000 & 1 & $2.0\times10^{-2}$ & 2.70 & $1.0\times10^{-3}$ & 1.46 \\
 (v) &70,000 & 1 & $8.8\times10^{-3}$ & 3.20 & $4.4\times10^{-4}$ & 1.73 \\
\hline
     &10,000 &0.1& $3.2\times10^{-1}$ & 12.2 & $2.2\times10^{-2}$ & 6.57 \\
(iii)&10,000 & 1 & $9.6\times10^{-2}$ & 1.21 & $7.2\times10^{-3}$ & 0.66 \\
     &10,000 &10 & $4.5\times10^{-2}$ & 0.12 & $2.3\times10^{-3}$ & 0.07 \\
\label{THR}\end{tabular}}\end{table}

We see from \Tab{THR} and \Fig{presults} clearly that as $\Teff$ increases 
for a given $g$, 
the radiative time steps (thick lines) decrease
whereas the acoustic time steps (thin lines) increase (both in the photosphere and deeper layers). 
We find that for stars with $\Teff < 10,000\K$,  $\delta t_{\rm s}$ gives the time step 
constraint, whereas for stars with $\Teff \gtrsim 10,000\K$, $\delta t_{\rm rad}$ 
is the more constraining one. 
Note that this segregation of the cold and hot stellar branches at $\Teff \approx 10,000\K$ 
seen in the $\delta t_{\rm rad}$ curves of \Fig{presults} is a consequence of our
opacity prescription given by \eq{Kramers}. In this equation,
we use $T_0 =  13,000\K$,  such that the stars 
hotter than this are described predominantly by Kramers opacity and stars cooler than this by
H$^-$ opacity.
We see from \Fig{presults} that for a given $g$, the $\delta t_{\rm s}$ and 
$\delta t_{\rm rad}$ curves intersect at a particular $\Teff$, 
which is typically $<10,000\K$; 
for e.g. the thin and thick red dashed lines in panel (a) intersect at $\sim 9000\K$. 
These stars are in fact the most economical 
to simulate numerically, as $\delta t_{\rm s} \approx \delta t_{\rm rad}$ and 
one does not need 
to worry about conflicting time steps in the problem. For hotter stars, however, we see 
that the $\delta t_{\rm s}$ and $\delta t_{\rm rad}$ diverge more and more 
away from each other, thus leading to a problem.
For a given $\Teff$, we see that $\delta t_{\rm s}$ always decreases with increasing $g$.
The $\delta t_{\rm rad}$ curves for a given $\Teff < 10,000\K$, on the other hand, seem to be 
nearly independent of $g$. However, as $\Teff$ increases beyond $10,000\K$, we find that 
the more massive stars have a shorter radiative time step. 
Interestingly, as $\Teff \rightarrow 10^5\K$, 
$\delta t_{\rm rad}$ again starts to become independent of $g$.  
Note that the time steps in the deeper layers of the star, as shown in \Fig{presults}(b) 
are much shorter than their photospheric counterparts, 
especially for hot massive stars. This also poses a numerical challenge if one 
wishes to simulate such a star all the way from the deeper layers up to the 
photosphere.
Figure~\ref{presults} can be thought to 
represent a Hertzsprung--Russell (HR) diagram for stars having different 
$g$ and $\Teff$.

Finally, we confirm that the stratification in the present
hydrostatic models agrees with the steady state solution
obtained in \Sec{sec_hotstar} using the {\sc Pencil Code}.
We do so by comparing the locations of the 
photosphere in \Figs{pdt_L256b2_axel1}{pdt_4panels}(d),
both of which correspond to the model having $\Teff \sim 70,000\K$.
We note from  \Fig{pdt_L256b2_axel1} that the computational 
domain ranges from $z=-30\Mm$ to $+30\Mm$, 
with the negative values representing the deeper layers of the star.
The corresponding
$\log_{10}p$ varies from $2.75$ to $5.47$ across the domain,
where $p$ is in cgs units.
On comparing with \Fig{pdt_4panels}(d), we see that this is well 
within the regime of static diffusion, where our numerical calculations 
are valid.
In \Fig{pdt_L256b2_axel1}, the photosphere 
($\tau=1$) is at $z=-5.8\Mm$ or $\log_{10}p=4.6$, which indeed is in 
excellent agreement 
with the location of the photosphere in \Fig{pdt_4panels}(d).
Also, the value of $\delta t_{\rm rad}$ at the photosphere 
agrees between \Figs{pdt_L256b2_axel1}{pdt_4panels}(d), 
being $\approx 10^{-2}\s$  
in both (keeping in mind that the time step in 
\Fig{pdt_L256b2_axel1} is given in  milliseconds).
We make a note here regarding the location of the 
minimum of the radiative time step in theoretical models 
versus numerical computation. We see from 
\Fig{pdt_4panels}(d) that $\min(\delta t_{\rm rad})$ occurs in 
the deeper layers at $\log_{10}p \approx 7.2$, which is 
in fact outside the computational domain of interest in \Fig{pdt_L256b2_axel1}. 
It is only for the cool stars, that the photosphere coincides with 
the location of $\min(\delta t_{\rm rad})$; see e.g. \Fig{pdt_4panels}(a).

\subsection{Simulation results for disc models}
\label{DNSdiscmodels}

In this section, we solve for the structure of an accretion disc 
around a white dwarf of mass $M_{\rm WD} = 1.1 M_\odot$. We implement the shearing box 
method in the {\sc Pencil code}, such that the disc is located at a distance of 
$r_{\rm disc}=10^{10} \cm$ from the central star, 
with $\varOmega=0.012 ~{\rm rad} \s^{-1}$ at this 
location. Thus, one rotation period at $r_{\rm disc}$ is 
$T_{\rm rot}=2\pi/\varOmega=0.52 \ks$.
In our earlier exploratory models,
we solved the disc models for both the
upper and lower disc plane, but in the subsequent models presented below, we
have restricted ourselves to solving the equations in just the upper disc plane
by assuming a symmetry condition at $z=0$.
These solutions are identical to our earlier ones and computationally more economic.

We consider one-dimensional solutions of \Eqss{dlnrho}{ngradI} for a
fixed value of the vertically integrated (surface) density
$\Sigma=\int_{-\infty}^\infty\rho\,\dd z$, 
with $\nu=5\times 10^{10}\cm^{2}\s^{-1}$ and $C_{\rm shock}=3$.
In our horizontally periodic domains with horizontal extent $L_x\times L_y$,
the total mass is $\Sigma L_x L_y$, which is conserved and therefore given
by the initial conditions.
Again, see appendix~\ref{ImprovedOuterBC} and 
appendix~\ref{hydrobc} for details about the boundary 
conditions on various variables.
We use $\Sigma=7\times10^{-6}\g\cm^{-3}\Mm$, and two values 
for ${\cal H}_0 = {\cal H}(z=0)$, namely,
${\cal H}_0=2\times10^{-6}\g\cm^{-3}\km^3\s^{-3}\Mm^{-1}$ (cold disc) and
${\cal H}_0=5\times10^{-5}\g\cm^{-3}\km^3\s^{-3}\Mm^{-1}$ (hot disc) 
and $z_{\rm heat}=1 \Mm$ for both cases.
For the initial isothermal stratification used here 
(as discussed in \Sec{AccretionDiscModel}), 
the pressure and density scale heights are identical, 
so $\Hp=c_{\rm s0}/\varOmega=\Hr$, but in general they are
somewhat different from each other; see appendix~\ref{ScaleHeights}. 
For the cold disc, we set $\Hp =0.42 \Mm$ ($c_{\rm s0} = 5 \km\s^{-1}$) and 
for the hot disc, $\Hp=2.5 \Mm$ ($c_{\rm s0} = 30 \km\s^{-1}$).
In cgs units, our values correspond to $\Sigma=7\times10^2\g\cm^{-2}$,
${\cal H}_0=20\g\cm^{-1}\s^{-3}$, and ${\cal H}_0=500\g\cm^{-1}\s^{-3}$, which
are appropriate values for discs in cataclysmic variables.
The corresponding accretion rates can be obtained from \eq{eq_heatfunc}, and 
are given by $\dot{M}=5.8\times 10^{13} \g \s^{-1}$ (cold disc) and 
$\dot{M}= 1.4 \times 10^{15}\g \s^{-1}$ (hot disc).
These were chosen based on preliminary calculations
of semi-analytically constructed models relevant to the regime where multi-valued
solutions of $\dot{M}$ (or equivalently ${\cal H}_0$ in our formalism) are 
possible for a given $\Sigma$. These solutions are governed by the 
hydrogen ionisation instability that lead 
to the so called disc instability model of cataclysmic variables; 
see \cite{Las01} for a review.
Note that for {\sc Pencil Code} simulations, it is convenient to measure lengths in units of
$\Mm$, speed in $\km\s^{-1}$, density in $\g\cm^{-3}$, time in $\ks$,
and temperature in Kelvin, which explains our
choice of units adopted in the results presented here.
Other choices would have been possible, too, although
working with $\Mm$ and $\km\s^{-1}$ is useful because
those are the units displayed in many of the diagrams.

\begin{figure*}[t!]\begin{center}
\includegraphics[width=\textwidth]{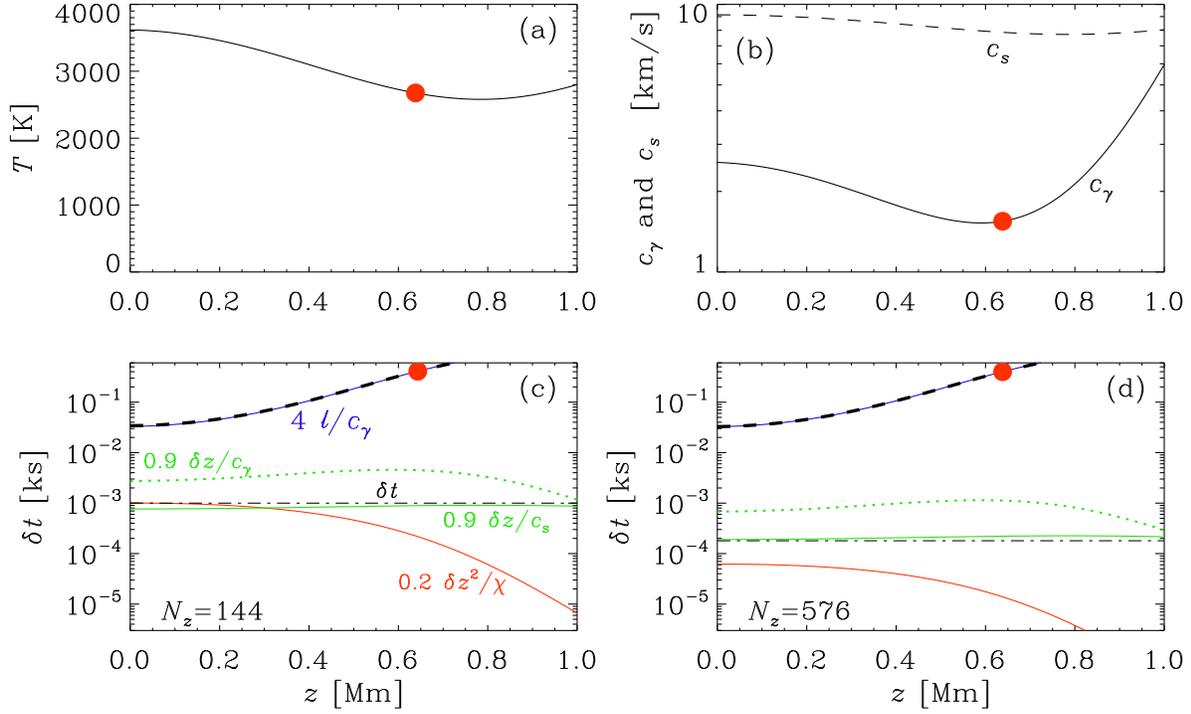}
\end{center}\caption[]{
Vertical profiles of $T$ (a), and $c_\gamma$ and $\cs$ (b). 
$\delta t_{\rm rad}^{\rm thin}$ (blue lines),
$\delta t_{\rm rad}^{\rm thick}$ (red lines),
their sum $\delta t_{\rm rad}$ (thick dashed lines),
$\delta t_{\rm s}$ (green solid lines),
$\delta t$ (black dot-dashed lines), 
and $\delta t_\gamma$ (green dotted lines)
for the cold disc model discussed in \Sec{DNSdiscmodels},
with $N_z=144$ (c) and $N_z=576$ (d).
All time steps are in kiloseconds
(colour online).}\label{pdt_tstep144Wa}\end{figure*}

We first discuss \Fig{pdt_tstep144Wa}, where we present the solutions 
for a cold accretion disc. In this case, 
$T$ varies from a photospheric
value of about $2600\K$ to about $3600\K$ in the midplane; see panel (a). 
Here we choose a vertical domain size of $L_z=1\Mm$.
The photosphere (i.e., $\tau=1$ surface), as denoted by 
the red dots in the figure, occurs at 
a depth of $0.65 \Mm$.
In panels (c) and (d), we plot the various time steps for two 
vertical grid resolutions $N_z=144$ and $576$, which were run for 
a total time of $3 \ks$ ($5.7~ T_{\rm rot}$) and 
$5 \ks$ ($9.7 ~T_{\rm rot}$), respectively.
According to our reasoning in \Sec{RadiativeCoolingConstraint},
the radiative time step should be limited by the sum of
$\delta t_{\rm rad}^{\rm thin}$ and $\delta t_{\rm rad}^{\rm thick}$. 
However, for the cold disc solution, the former is much larger. 
Here, $\delta t_{\rm rad}^{\rm thin}$ reaches values
of between $10^{-2}\ks$ in the midplane and increases to $10^{-1}\ks$
in the outer layers (for both $N_z=144$ and $576$ as it is independent of $\delta z$), 
and, hence it determines the radiative time step $\delta t_{\rm rad}$.
The maximum permissible empirically 
determined time step $\delta t$, on the other hand, is only around $10^{-3}\ks$ for 
$N_z=144$, and $1.8\times10^{-4}\ks$ for $N_z=576$.
It turns out that this value is entirely explained by the standard CFL
condition, where the time step is limited by the acoustic time step 
$\delta t_{\rm s}$; compare black dot-dashed and green solid lines in
figures~\ref{pdt_tstep144Wa}(c) and (d).
From panel (b) we see that $c_\gamma$ is well below $\cs$ throughout 
the computational domain. It varies only between $1.6\km\s^{-1}$ in the photosphere to 
about $2.6\km\s^{-1}$ in the midplane, and reaches $6\km\s^{-1}$ 
in the outermost parts well above the photosphere. 
The relevance of the ratio $\cs/c_\gamma$ will become more clear in 
\Sec{secMitigate}.

\begin{figure*}[t!]\begin{center}
\includegraphics[width=\textwidth]{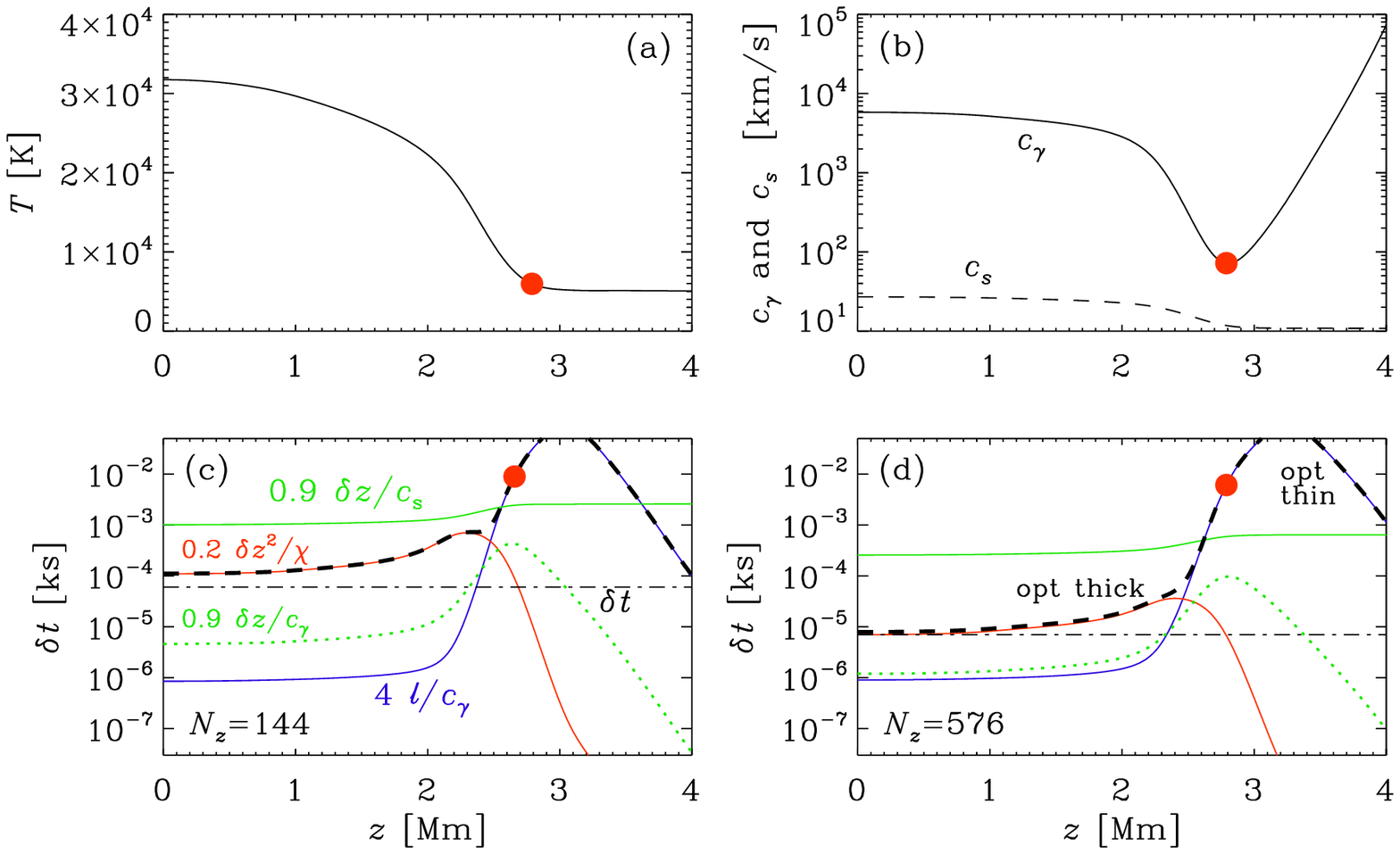}
\end{center}\caption[]{
Same as \Fig{pdt_tstep144Wa}, but for the hot disc model discussed in 
\Sec{DNSdiscmodels}.
All time steps are in kiloseconds
(colour online).}\label{pdt_tstep144e}\end{figure*}

In \Fig{pdt_tstep144e}, we present the hot disc solutions, 
where $T$ varies from a photospheric value
of $5000\K$ to about $32,000\K$ in the midplane.
We choose $L_z=4\Mm$ for this case, which extends from the midplane to a point
somewhat above the photosphere. 
The photosphere in this case occurs at a depth of $2.8 \Mm$.
We again show the 
results for $N_z=144$ and $576$, which were run for a total duration of $4\ks$ 
($7.7~T_{\rm rot}$) and $0.018\ks$ ($0.03~T_{\rm rot}$) respectively.
We find from panels (c) and (d) 
that the hot disc behaves somewhat similarly to 
the stellar surface layers discussed in \Sec{StellarSurfaceLayers}, where
$\delta t_{\rm rad}$ in the layers at some depth beneath
the photosphere is governed by $\delta t_{\rm rad}^{\rm thick}$ 
(solid red line), while in the outer layers by 
$\delta t_{\rm rad}^{\rm thin}$ (solid blue line). 
The minimum of the 
total radiative time step, however, occurs at the midplane, such that  
$\min(\delta t_{\rm rad}) \sim 10^{-4}\ks$ for $N_z=144$, and 
$\sim 7\times  10^{-6}\ks$ for $N_z=576$. 
This is very much consistent with the  maximum permissible empirical 
time step $\delta t$, which is 
$\sim 6\times10^{-5}\ks$ for $N_z=144$ and $\sim 7\times 10^{-6}\ks$ for $N_z=576$. 
Note that $\delta t_{\rm rad}^{\rm thick}$ is much more 
stringent in this case and limits the time
step; compare the red solid and black dot-dashed lines in \Fig{pdt_tstep144e}(d). 
Also, $\delta t_{\rm rad}^{\rm thick} \propto \delta z^2$, as can be seen 
from panels (c) and (d) of both \Figs{pdt_tstep144Wa}{pdt_tstep144e}.
The standard CFL condition is not relevant to explain $\delta t$ in this case as 
$\delta t_{\rm s} > \min(\delta t_{\rm rad})$.
Furthermore, $c_\gamma>c_s$ throughout the domain, reaching
about $7000\km\s^{-1}$ in the midplane and even $10^5\km\s^{-1}$
in the outer layers; see \Fig{pdt_tstep144e}(b).
We will return to the effect of this in \Sec{secMitigate}.

\begin{table}[b!]\caption{
Empirical time step $\delta t$ obtained from the one dimensional 
cold and hot disc simulations shown in 
\Figs{pdt_tstep144Wa}{pdt_tstep144e};
$\chi_0$ is given in $\Mm\km\s^{-1}$, $\csz$ in $\km\s^{-1}$,
and $\delta t$ in $\ks$, where $\chi_0$ and $\csz$ denote values
at the disc midplane.
The bold face values indicate the time step constraints obtained
for $C_{\rm rad}^{\rm thick}$ and $C_{\rm CFL}$.
}\vspace{12pt}\centerline{\begin{tabular}{lcc|ccc|ccc}
\hline
\hline
     &          &
     & \multicolumn{3}{|c}{$N=144$}
     & \multicolumn{3}{|c}{$N=576$} \\
     & $\chi_0$ & $\csz$
     & $\delta t$ & $\chi_0\delta t/\delta z^2$ & $\csz\delta t/\delta z$
     & $\delta t$ & $\chi_0\delta t/\delta z^2$ & $\csz\delta t/\delta z$ \\
\hline
Cold & $0.017$ & $9.1$
     & $1.0\times10^{-3}$ & 0.35 &      1.3
     & $1.8\times10^{-4}$ & 1.0  & {\bf0.95} \\
Hot  & $2.5$ & $27$
     &   $6\times10^{-5}$ & {\bf0.19} & 0.06
     &   $7\times10^{-6}$ &     0.35  & 0.03 \\
\label{Tsum}\end{tabular}}\end{table}

Finally, we discuss \Tab{Tsum}, where we summarise 
the empirically determined maximally
permissible time step $\delta t$ for the cold and hot disc 
models. We also compare with the numerically determined values of
$\chi_0\delta t/\delta z^2$ and $\csz\delta t/\delta z$, which will help us 
constrain the coefficients $C_{\rm rad}^{\rm thick}$ 
and $C_{\rm CFL}$, respectively
(the subscript 0 indicates the values of the respective quantities 
at the disc midplane). 
From the cold disc solutions, where the time step is 
constrained by $\delta t_{\rm s}$, we find that $\csz\delta t/\delta z=1.3$ 
for $N_z=144$, and $\csz\delta t/\delta z=0.95$ for $N_z=576$. 
These values are very close to the value of the standard 
Courant factor used in the {\sc Pencil Code}. Hence, we can 
conclude that $C_{\rm CFL}=0.95$, which is also consistent 
with the value of 0.9, which was adopted while plotting \Fig{pdt_tstep144Wa}. 
Note that ideally these values should be independent of 
resolution, but in practice they do seem to depend on it.
We hence adopt the smaller of the two values as the more restrictive 
constraint. From the hot disc solutions, 
where the time step is constrained by 
$\delta t_{\rm rad}^{\rm thick}$, we hope to constrain 
the coefficient $C_{\rm rad}^{\rm thick}$. First we find that for this case, 
$\csz\delta t/\delta z \ll 1$, 
which is contrary to our understanding of the CFL coefficient and, hence, 
we discard these values. Comparing the values of $\chi_0\delta t/\delta z^2$,
we conclude that $C_{\rm rad}^{\rm thick}=0.19$, since 
this is the smallest of the two resolutions. This value is also consistent with 
our choice of 0.2, which was used while plotting \Fig{pdt_tstep144e}.  
Note that neither the hot disc nor the cold disc models are suitable for 
determining the coefficient $C_{\rm rad}^{\rm thin}$, as 
$\delta t_{\rm rad}^{\rm thin}$ never constrains the time step in these cases.
However, the value of $C_{\rm rad}^{\rm thin}=4$ that we obtained from 
our earlier one-dimensional experiment and used for our plots, 
is indeed consistent with the simulation results reported in 
\Sec{sec_hotstar}; see also \Fig{pdt_L256b2_axel1}.

\begin{figure*}[t!]\begin{center}
\includegraphics[width=\textwidth]{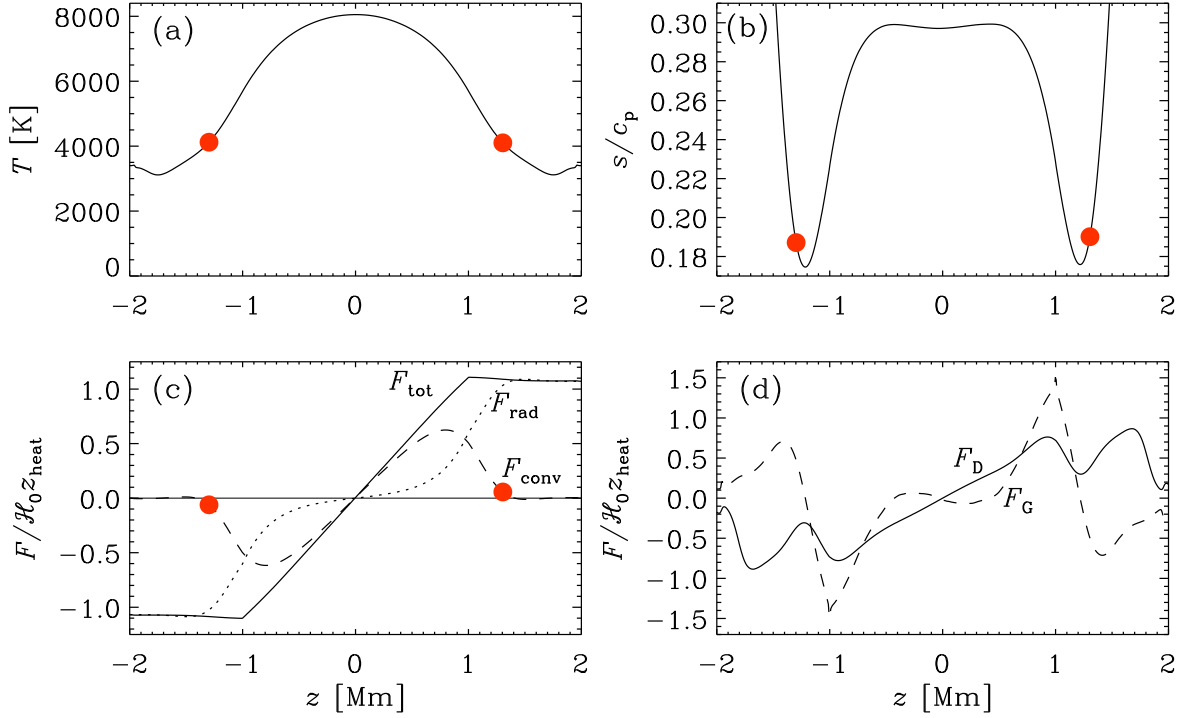}
\end{center}\caption[]{
Temperature (a) and specific entropy (b) along with the
various fluxes normalised by ${\cal H}_0 z_{\rm heat}$ (c) 
and (d), for a model with $\Sigma=7.2\times10^{-7} \g\cm^{-3}\Mm$ and
${\cal H}_0=10^{-5}\g\cm^{-3}\km^3\s^{-3}\Mm^{-1}$
(colour online).}\label{pflux_single2_comp}\end{figure*}

\subsection{Three-dimensional simulations of discs with vertical heating profile}
\label{3DdiscDNS}

In \Sec{AccretionDiscModel}, we introduced the heating profile
${\cal H}(z)$ and stated that the discontinuities at $z=\pm z_{\rm heat}$
do not cause any artefacts in the thermodynamic variables
such as temperature and specific entropy.
This is demonstrated in \Fig{pflux_single2_comp}, where we show the
results of a three-dimensional simulation (of both the gas and the
radiation field) for the same white dwarf-accretion disc 
system as in \Sec{DNSdiscmodels}, but with an intermediate heating source
${\cal H}_0=10^{-5}\g\cm^{-3}\km^3\s^{-3}\Mm^{-1}$ (or equivalently, 
$\dot{M} = 2.9 \times 10^{14} \g \s^{-1}$) in a cubic domain
$[L_x, L_y, L_z]$ of size $(4\Mm)^3$, covering both the lower and upper
disc planes, using $576^3$ mesh points, $z_{\rm heat}=1\Mm$, 
$\nu=5\times10^9 \cm^2 \s^{-1}$ and $C_{\rm shock}=0.5$.
Again, see appendices~\ref{ImprovedOuterBC} and 
\ref{hydrobc} for details about the boundary 
conditions on various variables.
The 3D simulation was run for a total time of $27.5\ks$ or $52.8~T_{\rm rot}$.

The temperature reaches a maximum of about $8000\K$ at $z=0$ and has an
approximately flat profile away from the lower and upper photospheres 
(i.e., at $\pm 1.3 \Mm$).
It turns out that the $z$ dependence of temperature profile is not
perfectly flat.
This can probably be ascribed to the long thermal adjustment time in
this system, which we can estimate as follows.
The Kelvin--Helmholtz timescale is given by $\tau_{\rm KH}=E_{\rm th}/L$,
where $E_{\rm th}=\int\rho\cv T\,\dd V$ is the internal energy,
$L=2F_\infty L_x L_y=2{\cal H}V$ is the luminosity for the losses on
both photospheres, $V=L_x L_y L_z$ is the volume of the domain, 
and $F_\infty$ is the value of the total flux in the photosphere.
Using $E_{\rm th}=\Sigma L_x L_y\overline{\cs^2}/[\gamma(\gamma-1)]$ with
$\overline{\cs^2}\equiv\bra{\rho\cs^2}/\bra{\rho}\approx(13\km/\s)^2$,
and $[\gamma(\gamma-1)]^{-1}=0.9$ for $\gamma=5/3$, we have
$\tau_{\rm KH}=0.45\,\overline{\cs^2}\Sigma/{\cal H}_0L_z=50\ks$,
which is about three times longer than the duration of our simulation.

The specific entropy, defined here as $s/\cp=\ln(T/p^{\nabad})$, has a negative slope,
$\dd s/\dd\ln z<0$, corresponding to a Schwarzschild-unstable stratification.
This should lead to instability and hence to turbulent convection.
This is indeed the case as seen in \Fig{pflux_single2_comp}(c), where we plot
the mean energy fluxes averaged over a time span when the system is
in an approximately steady state between $t=15\ks$ and $20\ks$,
which is still subject to slow thermodynamic adjustments on
the longer timescale $\tau_{\rm KH}=50\ks$.
In particular, we plot $F_{\rm rad}$ together with $F_{\rm conv}$
and $F_{\rm tot}=F_{\rm rad}+F_{\rm conv}$, where
$F_{\rm conv}=F_{\rm enth}+F_{\rm kin}$, with
\EQ
F_{\rm enth}=\overline{\rho u_z\cp T},\qquad
F_{\rm kin}=\overline{\rho u_z\uu^2}/2,
\EN
being the enthalpy and kinetic energy fluxes, and overbars denote from
now on horizontal $xy$ averages.

Note that in \Fig{pflux_single2_comp}, $F_{\rm tot}$ varies approximately
linearly in $|z|<z_{\rm heat}$ and then reaches a plateau with
$F_{\rm tot}=\pm F_{\rm tot}^\infty$ on both ends.
However, $F_{\rm tot}^\infty/{\cal H}_0 z_{\rm heat}\approx1.2$
exceeds the expected values of $\pm1$, which is, again, indicative
of the simulation still not being in thermal equilibrium.
The radiative flux is small near the midplane and energy is mostly
carried by convection.
The kinetic energy is not plotted, but it is about 10\% of the
convective flux and directed opposite to it, i.e., inward.
This is a well known consequence of the up-down asymmetry of
compressible convection \citep{HTM84}.

We see that in the inner parts of the disc, i.e., for $|z|<0.5\Mm$,
most of the energy is carried by convection.
This is curious in view of the fact that the stratification in those
parts is close to adiabatic.
Therefore, the standard mixing length prescription of the convective
flux being carried by a gradient term, $F_{\rm enth}\approx F_{\rm G}$
with \citep[e.g.][]{Rue89}
\EQ
\FF_{\rm G}=-\chi_{\rm turb}\meanrho\meanT\nab\means,
\label{FG}
\EN
where $\chi_{\rm turb}\approx\urms\Hp/3$, cannot hold.
As argued in \cite{Bra16}, the reason for this is that there is another
important term, the Deardorff term, resulting from entropy fluctuations
\citep{Dea66,Dea72},
\EQ
\FF_{\rm D}=-\tau\overline{s'^2}\grav/\cp,
\EN
where we have ignored the possibility of factors of the order of unity.
Likewise, in \eq{FG}, there could also be such factors, so we cannot
expect perfect agreement between $F_{\rm conv}$ and the contributions
$F_{\rm G}$ and $F_{\rm D}$ shown in \Fig{pflux_single2_comp}(c).
We do see, however, that $F_{\rm D}$ increases approximately linearly
in the bulk of the disc ($|z|<0.5\Mm$), while $F_{\rm G}$ does not even have 
the correct sign in order to explain $F_{\rm conv}$.
This is strong evidence that convection in such discs must be described
by the Deardorff term.
Similar results have previously only been found from stellar convection
simulations \citep{Kapy_etal17,KVKB18}.

\begin{figure*}[t!]\begin{center}
\includegraphics[width=\textwidth]{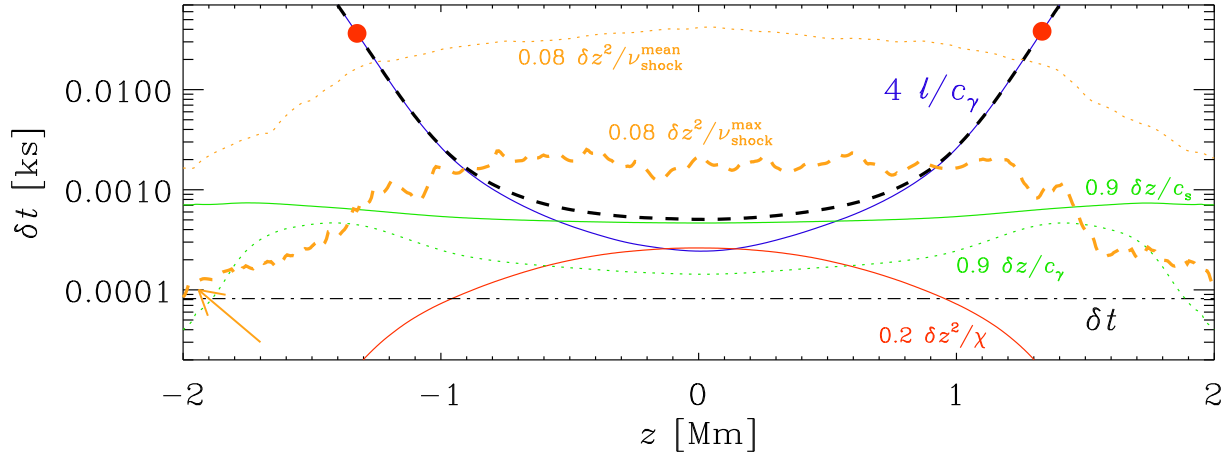}
\end{center}\caption[]{
$z$ dependence of the various time steps for the three-dimensional
convective accretion disc simulation with
$\Sigma=7\times10^{-6}\g\cm^{-3}\Mm$ and
${\cal H}_0=10^{-5}\g\cm^{-3}\km^3\s^{-3}\Mm^{-1}$.
$\delta t_{\rm rad}^{\rm thin}$ (blue solid line), $\delta t_{\rm rad}^{\rm thick}$ 
(red solid line), their sum $\delta t_{\rm rad}$ (black dashed line), 
$\delta t_{\rm s}$ (green solid line),
$\delta t_\gamma$ (green dotted line), the time step due to 
the maximum shock viscosity 
($0.08 \delta z^2/\nu^{\rm max}_{\rm shock}$; orange dashed line), 
the time step due to mean shock viscosity
($0.08 \delta z^2/\nu^{\rm mean}_{\rm shock}$; orange dotted line),
and the empirical time step $\delta t$ (black dot-dashed line).
All time steps are in kiloseconds.
The arrow indicates the location from where the limiting time step
constraint originates
(colour online).}\label{ppdt_Strat_H1.0e-05_z20_nu5em4a4}\end{figure*}

Returning to the topic of this paper, we now investigate the various
times step constraints for this model.
In \Fig{ppdt_Strat_H1.0e-05_z20_nu5em4a4}, similar to our earlier plots,
we show the various times steps, but now we also plot the time step
constraint from viscosity, particularly the one from shock viscosity.
Note that the empirical time step $\delta t$ in this case is not 
fixed at the beginning of the simulation 
(as done in figures \ref{pdt_L256b2_axel1}, \ref{pdt_tstep144Wa} and 
\ref{pdt_tstep144e}), but is allowed to evolve.
It turns out that the minimum of radiative and acoustic time steps are
approximately equal ($\approx5\times10^{-4}\ks$) and occur in the disc
midplane ($z=0$).
However, the most severe time step constraint comes, in this case, from
the outer layers above the photosphere, where the shock viscosity is large 
and hence the corresponding time step is small (green dashed line). 
This is in accord with the empirical time step
$\delta t \approx 8 \times 10^{-5} \ks$ attained
at the end of the the simulation.

\begin{figure*}[t!]\begin{center}
\includegraphics[width=\textwidth]{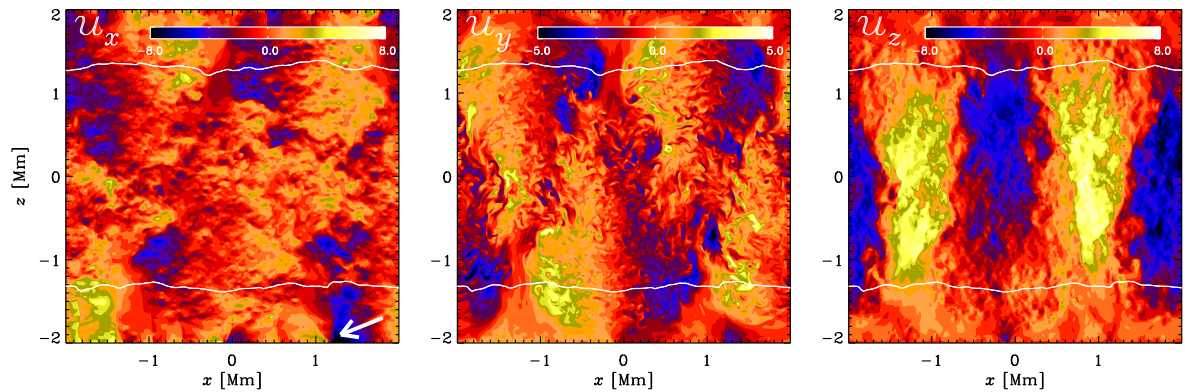}
\end{center}\caption[]{
$xz$ cross sections of $u_x$, $u_y$, and $u_z$ (in $\km\s^{-1}$) for 
the three-dimensional convective accretion disc simulation with
$\Sigma=7\times10^{-6}\g\cm^{-3}\Mm$ and
${\cal H}_0=10^{-5}\g\cm^{-3}\km^3\s^{-3}\Mm^{-1}$.
The white lines show the $\tau=1$ surfaces and 
the arrow points to the strongest shock near $z=-2\Mm$
(colour online).}\label{pslice_Strat_H1.0e-05_z20_nu5em4a4}\end{figure*}

To appreciate the reason for such shocks to occur, we now look at the $xz$
cross sections of $u_x$, $u_y$, and $u_z$ shown in \Fig{pslice_Strat_H1.0e-05_z20_nu5em4a4}.
We see that the flow possesses two major pairs of up- and down-draughts.
These cross sections were taken at $y=0$, but different cross sections for
different values of $y$ look qualitatively similar, indicating that the
large-scale structure is independent of $y$.
Away from the both photospheres, however, significant velocities are still
possible, which can lead to the formation of shocks in those parts.
These are best seen in the image of $u_x$; 
see first panel of \Fig{pslice_Strat_H1.0e-05_z20_nu5em4a4}.
The arrow points to the strongest shock near $z=-2\Mm$, where the
local Mach number, i.e., the ratio of velocity to sound speed,
reaches values of around 1.3.

\section{Approaches to mitigating the radiative time step problem}
\label{secMitigate}

There appears to be a general problem with hot stars and accretion discs
where the radiative time step is much shorter than the acoustic one.
This is either because
\EQ
4\,\ell/c_\gamma\ll0.9\,\delta z/\cs
\qquad\mbox{(problem~A)},
\label{eqproblemA}
\EN
as in \Sec{sec_hotstar}, or because
\EQ
0.2\,\delta z^2/\chi\equiv0.6\,\delta z^2/c_\gamma\ell\ll0.9\,\delta z/\cs
\qquad\mbox{(problem~B)},
\EN
as in \Sec{DNSdiscmodels}.
In both cases, the problem is caused by the smallness of
the factor $\Bo=16\cs/c_\gamma$
compared with the value of $\delta z/4\ell$ in problem~A or the
value of $\ell/0.6\delta z$ in problem~B.
In problem~A, $\cs/c_\gamma$ becomes as small as $10^{-6}$,
as seen by comparing the green solid and 
dotted lines at the outer edge of the computational domain
in \Fig{pdt_L256b2_axel1}(b).
While in problem~B, $\cs/c_\gamma$ drops to
$10^{-4}$ at the outer edge of the disc, as seen in \Fig{pdt_tstep144e}(b).
For problem~A, the photospheric value of $\delta z/4\ell$ is 
about 0.006, as calculated from the model discussed in \Fig{pdt_L256b2_axel1}. While for 
problem~B, the photospheric value of $\ell/0.6\delta z$ is about $16$, as 
calculated from the model discussed in \Fig{pdt_tstep144e}(d). Note that 
these are both quite large compared to the respective $\cs/c_\gamma$ values.

To discuss ways of mitigating the time step problem, we must distinguish
the two cases A and B.
For problem~B, the difficulty appears through the optically thick constraint
in the part where convection would develop in the three-dimensional simulations,
as discussed in \Sec{3DdiscDNS}; see also \Fig{pslice_Strat_H1.0e-05_z20_nu5em4a4}.
Thus, radiation should not be physically important and it would be
unreasonable to spend much computational resources on this.
One may therefore adopt the approach of artificially increasing the opacity in that
part, so that $\chi=c_\gamma\ell/3=c_\gamma/3\kappa\rho$ becomes smaller
and, as a consequence, the optically thick time step becomes longer.
This would increase the fractional convective flux, but not the
total flux.

For problem~A, on the other hand, it is important to maintain a large
radiation pressure in order to study photoconvection and, hence, 
$\kappa F_{\rm rad}/cg$ should be close to unity.
Here, however, increasing $\kappa$ would be counterproductive, because
the relevant time step is $4\,\ell/c_\gamma$, which would become
even smaller as $l \propto 1/\kappa$.
Thus, we have the problem that, on the one hand, $\kappa F_{\rm rad}/cg$
should be close to unity and certainly not be too small, and on the other
hand, $4\, \ell \,\cs/(0.9 \,\delta z \,c_\gamma)$ should also be close
to unity so that the acoustic and optically thin radiative time steps
are close together; see \Eq{eqproblemA}.
Combining these two constraints, we want to ensure that the
\EQ
\mbox{product of two constraints}
=\underbrace{{4\,\ell\over0.9\,\delta z}\,
{\cs\over c_\gamma}}_{\mbox{\scriptsize time step}}\,
\underbrace{\kappa F_{\rm rad}\over cg}_{\mbox{\scriptsize photoconv}}
\EN
is as close to unity as possible because we want
$\kappa F_{\rm rad}/cg \gg 1$.
Using $F_{\rm rad}\approx\sigmaSB T^4$, $\delta z\approx0.05\Hp$,
and $\kappa\rho\ell=1$, we find for the
\EQ
\mbox{product of two constraints}
={4.8\, \cs \over 0.05\times16\, \nabad\, c}\approx15\,\cs/c\approx\Bo.
\label{constraintproduct}
\EN
Thus, it is clear that the best chance of studying photoconvection
in hot massive stars is in the relativistic regime, because then 
the right hand side of \Eq{constraintproduct} is close to unity.
In that case, however, one can no longer neglect the time
dependence of the radiation field.
Alternatively, of course, one may just take a solar-type model, where
the radiative and acoustic time steps are already close together, but then decrease
$c$ artificially, in order to boost the radiative pressure and reduce 
computational costs.
This approach is similar to the reduced speed of light approximation (RSLA) 
originally proposed by \cite{GA01}. It has been employed, for e.g., by \cite{SO13}, 
who adopt a particular closure relation and solve the radiation momentum equation 
semi-explicitly, in a module for the Athena code \citep{SGTHS08}. 
They found that the RSLA method is best applicable
for systems with moderate optical depth ($\lesssim 10$), such as several 
star formation regimes in galactic discs.

At the end of \Sec{RadiativeCoolingConstraint}, we mentioned the
possibility that also a strong radiation pressure could in principle
restrict the time step.
Such a situation may arise, for example, in the deep interior 
of a super-massive star; see \Fig{pdt_4panels}.
A relevant application of the RSLA is therefore the study of such
a constraint.
We therefore now reconsider the simulation of the hot star of
\Sec{sec_hotstar}, where the radiation pressure was retained
in the momentum equation \eq{DuDt}.
We artificially lower the value of $c$ in the term
$g_{\rm rad}=\kappa F_{\rm rad}/c$ of this equation
to make it more pronounced.
We find that it is only when we lower it from $c=3\times10^5\km\s^{-1}$
to about $0.02\km\s^{-1}$ that this term begins to restrict the time step.

\begin{table}[b!]\caption{
Values of $\delta t\,\sqrt{g_{\rm rad}/\delta z}$ for the 
maximum permissible time step 
$\delta t$ for different values of $c$ and $g$.
}\vspace{12pt}\centerline{\begin{tabular}{c|ccccc}
\hline
\hline
$c$ [$\km\s^{-1}$] & $3\times10^{5}$ & $2\times10^{-2}$ & $10^{-2}$ & $10^{-3}$ \\
\hline
$g$ [$\km^2\s^{-2}\Mm^{-1}$] & $2.73\times10^{2}$ & $2.15\times10^{9}$ & $4.31\times10^{9}$ & $4.32\times10^{10}$ \\
$\delta t$ [$\ks$] & $4.7\times10^{-6}$ & $4.8\times10^{-6}$ & $3.5\times10^{-6}$ & $9.5\times10^{-7}$ \\
$\delta t\,\sqrt{g_{\rm rad}/\delta z}$ & $10^{-4}$ & $0.46$ & $0.47$ & $0.41$ 
\label{Tradpres}\end{tabular}}\end{table}

In \Tab{Tradpres}, we list such models, where we lower the value of $c$
and increase $g$ such that $g\approx g_{\rm rad}$ so that the model
remains in Eddington equilibrium, i.e., $g_{\rm eff}\to0$ in \Eq{eq_geff}.
Note that all the other parameters of the simulation remain the same as mentioned in 
\Sec{sec_hotstar}.
The time needed to accelerate a parcel of gas over a distance $\delta z$
is $\sqrt{2\delta z/g_{\rm rad}}$.
We can therefore formulate a new time step constraint by generalising
\eq{dtfull} to
\begin{equation}
\delta t=\min\left(..., C_{\rm rad}^{\rm pres}\,
\delta z^{1/2}/g_{\rm rad}^{1/2}\right),
\label{dtfull2}
\end{equation}
with a new coefficient $C_{\rm rad}^{\rm pres}$ and the ellipsis
indicating the other terms of \Eq{dtfull}.
This additional constraint is qualitatively different from any of our
earlier constraints in that it is proportional to $\delta z^{1/2}$.
The results presented in \Tab{Tradpres} suggest that
$C_{\rm rad}^{\rm pres}\approx0.4$.
This new finding illustrates the potential usefulness of applying
the RSLA.

By changing or modifying the physical problem, one is clearly never
able to solve the actual problem, but, under certain circumstances,
one may be able to address relevant aspects of the problem.
An example of this type of approach is the work of \cite{KMCWB13,KGOKB19},
who study astrophysically relevant aspects of rotating convection by
firstly increasing the radiative flux to bring the Kelvin-Helmholtz
timescale closer to other timescales in the problem, and secondly
increasing the angular velocity such that the rotational
influence on the flow is in the relevant regime.
How useful such approaches are cannot easily be answered and may
ultimately need to await a full solution to the problem.
Another example is the study of high Rayleigh and Reynolds number flows,
where we will never be able to solve the astrophysically relevant regime
using brute force approach.

\section{Conclusions}

Our work has demonstrated quantitatively how the maximally permissible
time step varies with $\delta z^2/\chi D$ in the optically thick regime
and with $\ell/c_\gamma$, independent of $\delta z$, in the optically
thin regime.
In particular, we have shown that the radiative time step is governed
by the sum of the optically thick and thin constraints, and not
by the smaller of the two, as one might have naively expected by analogy
with other time step constraints entering the problem.
This agrees with the work of \cite{FSL12}, except that our studies
identify the existence of two independent 
coefficients in front of the terms for the optically thin and thick regimes.
It should be noted, however, that the large difference between
the coefficients $C_{\rm rad}^{\rm thick}$ and $C_{\rm rad}^{\rm thin}$
could have been alleviated by working with the Nyquist wavenumber instead
of the inverse mesh spacing.
This amounts to including a $\pi^2$ factor in the
definition of $C_{\rm rad}^{\rm thick}$; see our discussion in
\Sec{RadiativeCoolingConstraint}.
However, to facilitate comparison with earlier work, we have refrained
from redefining this coefficient.

The fact that we find $C_{\rm rad}^{\rm thick}\approx0.2$ and
$C_{\rm rad}^{\rm thin}\approx4$ over a broad range 
of different physical circumstances supports our expectation that
these values are universal.
While this is to be expected, we should nevertheless keep an open mind
and continue verifying this in new and as yet unexplored regimes whenever
possible.

We have also demonstrated that a radiative time step constraint of the form
$\delta z/c_\gamma$, analogous to the Courant condition as given 
by \eq{eq_davis12}, is not, in general,
justified; see especially our \Figs{pdt_L256b2_axel1}{pdt_tstep144e}.
We thank Matthias Steffen (private communication) for pointing out to us
that such a relation could be justified in principle 
by noting that \eq{US66expr} predicts
maximum cooling when $\ell=\sqrt{3}/k_{\rm Ny}\approx0.55\,\delta z$, so
that the cooling time becomes $2\ell/c_\gamma\approx1.1\delta z/c_\gamma$.
The simulation presented in \Fig{pdt_L256b2_axel1} has demonstrated,
however, that this would result in a time step constraint that would be about
1000 times shorter (see minimum of green dotted line in \Fig{pdt_L256b2_axel1})
than the actual maximum permissible time step
(see black dot-dashed line) required for stability.
Thus, $\delta z/c_\gamma$ does not appear to be a useful indicator of
the maximum permissible time step.

Incidently, \Figp{pdt_4panels}{a} shows that, in the proximity of the
photosphere, the local value of $\delta z/c_\gamma$ is indeed close to
the actual maximum permissible time step of about $10^4\s$.
However, already in \Figp{pdt_4panels}{b} this is no longer the case,
because the black dashed line attains its minimum
well below the photosphere, where $\log_{10}p\approx8$.
Furthermore, in \Figp{pdt_4panels}{a}, the expected time step constraint 
is apparently not given by the local
minimum of $\delta z/c_\gamma$, but rather by the local maximum
of $\delta z/c_\gamma$.
This is here because $c_\gamma$ has a minimum close to the photosphere.
Deeper down it increases because $T$ increases, and higher up it also
increases because $\rho$ drops rapidly.

By covering stellar surface models in the HR diagram, we get a
comprehensive understanding of which of the two constraints decide about
the limiting time step for a large range of different circumstances.
This showed that the most severe constraint on the radiative time step
occurs for larger values of $g$ and $\Teff$. This corresponds to the
lower left corner of the HR diagram, recalling that in theoretical HR diagrams,
increasing luminosity is replaced by decreasing surface gravity;
see, e.g., \cite{TACNS13}.
We have also seen that, for the cool stars with $\Teff \lesssim 5000\K$
the location of the minimum radiative time step coincides with the
location of the photosphere. This is
not true for hotter stars, where the minimum radiative time step 
occurs in the deeper layers. For accretion discs (both hot and cold), 
on the other hand, the shortest radiative time step tends to occur
in the midplane.

The examples presented in this paper highlight some of the difficulties
in dealing with global simulations by covering regimes where different time
step constraints prevail.
It is clear that the optimal approach would be one where different regions
in space would not only have different spatial resolutions, but
also different time steps.
This would save resources that can at the same time be used to speed
up the calculation in regions that require shorter time steps.
A code satisfying such requirements is the {\tt DISPATCH} code \citep{NRPK18}.
Nevertheless, the time step constraints discussed in the present paper
should apply to such codes just as well.

For the three dimensional accretion disc model presented here, the temperatures are
moderate, and the acoustic and radiative time step constraints are about
equally short.
This happens in the midplane.
In the outer parts above the two photospheres, shocks become
important.
The time step constraint resulting from the shock viscosity is here about
equally severe as the acoustic and radiative ones in the disc midplane.
In such a situation, one might not gain much by using an implicit scheme for radiation,
or by adopting a code that treats different regions in space with
different times steps.
However, it is important to monitor the various times step constraints
carefully and try to stay close to physical regimes in which the
different constraints are not vastly different from each other. In this way,
the simulation can utilise existing resources in an optimal way.

\section*{Acknowledgements}

This paper is dedicated to Ed Spiegel.
If it was not for the radiative time step problem, the
Brandenburg and Spiegel (2006) paper would have been published by now!
We thank Matthias Steffen for sharing with us his experience and knowledge
regarding the time step constraint in stellar convection simulations.
We are also grateful to the three referees for their thoughtful comments.
AB also acknowledges Fazeleh (Sepideh) Khajenabi for her work on the
radiative time step problem while visiting Nordita in the spring of 2010.
UD thanks Akshay Bhatnagar for useful discussions.
This research was supported in part by the National Science Foundation
under the Astronomy and Astrophysics Grants Program (grant 1615100),
and the University of Colorado through its support of the George Ellery
Hale visiting faculty appointment.
Simulations presented in this work have been performed with computing
resources provided by the Swedish National Allocations Committee at
the Center for Parallel Computers at the Royal Institute of Technology
in Stockholm.
The source code used for the simulations of this study,
the {\sc Pencil Code}, is freely available on
\url{https://github.com/pencil-code/}.
The DOI of the code is \url{http://doi.org/10.5281/zenodo.2315093}.
The setups of runs and corresponding data are freely available on
\url{https://www.nordita.org/~brandenb/projects/tstep/}.


\appendix

\section{Radiative boundary conditions}
\label{ImprovedOuterBC}

In \Sec{RadiationTransport}, we stated that the assumption of zero
incoming intensity is not accurate when we want to reproduce the analytic
solution for an infinitely extended layer, where the gas beyond the
simulated boundary does contribute to producing incoming radiation.
To take this into account, we assume that $F_{\rm rad}$ is known and
that the system is in radiative equilibrium.
In that case, and with just two rays, $I_\pm=I(\xx,t,\pm\zzz)$, we have
\EQ
S=J=(I_++I_-)/2\quad\mbox{and}\quad
F_{\rm rad}/4\pi=(I_+-I_-)/2,
\EN
so $I_\pm=(S\pm F_{\rm rad}/4\pi)/2$.
In the {\sc Pencil Code}, this boundary condition is invoked by stating
the symbolic name {\tt bc\_rad=`p',`p',`S+F:S-F'}. The three entries 
correspond to the three 
directions $(x,y,z)$ respectively, {\tt `p'} represents periodic boundary 
condition, {\tt S+F} applies to the lower $z$ boundary and 
{\tt S-F} applies to the upper $z$ boundary (and are thus separated by 
a colon). Note that these are the 
boundary conditions used for the hot stellar model in \Fig{pdt_L256b2_axel1}.

The radiative boundary conditions used for the 1D accretion disc 
models in \Figs{pdt_tstep144Wa}{pdt_tstep144e} 
are {\tt bc\_rad=`p',`p',`S:0'}, where 
{\tt 0} represents zero value in ghost zones and 
free value on the upper $z$ boundary. The conditions used for the 
three dimensional accretion 
disc model in \Sec{3DdiscDNS} are {\tt bc\_rad=`p',`p',`0'}.

\section{Different formulations of the energy equation}
\label{SpiegelVeronis}

The energy equation can be formulated in a number of equivalent forms.
\EEq{DsDt} illustrates that heating and cooling only affect the specific
entropy, while \Eq{opt_thickthin} illustrates that temperature change
is governed by the specific heat at constant {\em pressure} when
$\DD p/\DD t$ can be neglected, which is when sound waves equilibrate
pressure fluctuations.
Another formulation of the right-hand side of \Eq{DsDt} is
\EQ
\rho\cv{\DD T\over\DD t}-p{\DD\ln\rho\over\DD t}
={\cal H} - \nab \bm \cdot\FF_{\rm rad}+\ttau\bm{:}\nab\UU,
\label{DTDt_with_Dlnrho}
\EN
which shows that temperature changes depend on the specific heat at
constant {\em volume} when $\DD\ln\rho/\DD t$ can be neglected.
The formulation in \eq{opt_thickthin} has been favoured by \cite{SV60}
to show that in the Boussinesq approximation, where $\nab \bm \cdot\uu=0$,
the relation $\DD\ln\rho/\DD t=-\nab \bm \cdot\uu$ is no longer invoked and
$\DD\ln\rho/\DD t$ cannot be neglected.
It is this fact that also motivates the presence of the $\gamma=\cp/\cv$
factor in the definition of the Prandtl number $\Pr=\nu/\gamma\chi$.

\section{Boundary conditions for hydrodynamic variables}
\label{hydrobc}

The purpose of this appendix is to discuss numerical details regarding
the models presented in \Sec{sec_hotstar}, \Sec{DNSdiscmodels} 
and \Sec{3DdiscDNS}.

The independent variables solved for the hot stellar surface model in 
\Fig{pdt_L256b2_axel1} are
$u_x$, $u_y$, $u_z$, $\ln \rho$, $\ln T$ and $\nu_{\rm shock}$. 
The boundary conditions supplied for these 6 variables along the 
$z$-direction is given in {\sc Pencil Code} notation as, 
{\tt bcz=`s',`s',`a',`e2',`e2',`s'}; where {\tt `s'} implies 
symmetry or vanishing first derivative, {\tt `a'} implies antisymmetry 
or vanishing value, and {\tt `e2'} implies extrapolated value.

The independent variables solved for the 1D accretion disc models
presented in \Figs{pdt_tstep144Wa}{pdt_tstep144e} are, $u_x$, $u_y$,
$u_z$, $\ln \rho$, $s$ and $\nu_{\rm shock}$.
The boundary conditions supplied for these 6 variables along the
$z$-direction is given by
{\tt bcz=`s',`s',`a',`s:a2',`s:a2',`s'}; where {\tt `a2'} implies 
antisymmetry with a vanishing second derivative. 

The independent variables solved for the 3D accretion disc model presented in
figures~\ref{pflux_single2_comp}--\ref{pslice_Strat_H1.0e-05_z20_nu5em4a4} 
are, $u_x$, $u_y$, $u_z$, $\ln \rho$, $s$ and 
$\nu_{\rm shock}$. The boundary conditions supplied for these 6 variables along the
$z$-direction is given by {\tt bcz=`s',`s',`a',`a2',`a2',`s'}, while periodic 
boundary conditions are used along $x$ and $y$.

\section{Relation between pressure and density scale heights}
\label{ScaleHeights}

As stated in \Sec{DNSdiscmodels}, $\Hp=\Hr$ for an isothermal
stratification.
For an isentropic stratification, we have $\gamma \Hp=\Hr$.
In terms of the double-logarithmic temperature gradient,
$\nabla=\dd\ln T/\dd\ln p$, the general relation is given by
\EQ
\nabla-\nabad={\dd(s/\cp)\over\dd\ln p}
={1\over\gamma}-{\dd\ln\rho\over\dd\ln p}
={1\over\gamma}-{\Hp\over\Hr}.
\EN
Thus, since $\nabad=1-1/\gamma$, we have
\EQ
\Hp/\Hr=\nabla-1,
\EN
which is independent of $\gamma$.
In the absence of convection, and for simple power law opacities
given by \eq{Kramers} with single exponents $a$ and $b$,
$\nabla=1/(1+n)$ depends
on the polytropic index $n=(3-b)/(1+a)$ \citep{BB14}.
In \Tab{ScaleHeightRatio}, some examples are listed.

\begin{table}[h!]\caption{
Examples of $\Hp/\Hr$ for different types of stratification.
}\vspace{12pt}\centerline{\begin{tabular}{lccclll}
      & $\gamma$ &$a$& $ b  $ & $n$ &$\nabla$&$\Hp/\Hr$ \\
\hline
isentropic & 5/3 & 1 & $ 0  $ & 1.5  & 0.4   & 0.6  \\
elect scattering & 5/3 & 0 & $ 0  $ & 3    & 0.25  & 0.75  \\
Kramers opacity & 5/3 & 1 & $-3.5$ & 3.25 & 0.235 & 0.765 \\
           & 5/3 & 0 & arbitr &$\infty$& 0   & 1   \\
isothermal &  1  & 0 & arbitr &$\infty$& 0   & 1 (indep of $\gamma$) \\
\label{ScaleHeightRatio}\end{tabular}}\end{table}

\label{lastpage}
\end{document}